\documentclass[lettersize,journal]{IEEEtran}

\usepackage[T1]{fontenc}

\usepackage{cite}

\usepackage{amsmath,amssymb}

\usepackage{bm}
\usepackage{array,enumitem}
\usepackage{booktabs}
\usepackage{siunitx}
\usepackage{dsfont}
\usepackage{subfigure}
\usepackage{bbm}
\usepackage[dvipsnames,table]{xcolor}
\usepackage{tikz,pgfplots}
\usetikzlibrary{pgfplots.groupplots}
\usetikzlibrary{patterns,shapes,fit}
\pgfplotsset{ every non boxed x axis/.append style={x axis line style=-},
	every non boxed y axis/.append style={y axis line style=-}}
\pgfplotsset{minor grid style={dotted,gray!25}}
\pgfplotsset{major grid style={dashed,gray!25}}
\usepackage[toc,acronym,nonumberlist,nopostdot]{glossaries}
\makeindex
\makenoidxglossaries

\renewcommand{\sf}[1]{{\mathsf{#1}}}

\newcommand{\E}[2][]{\mathbb{E}_{#1}\!\left[#2\right]}
\renewcommand{\P}[1]{\mathbb{P}\!\left[#1\right]}
\renewcommand{\vec}{\bm}

\newcommand{\norm}[2][2]{\left\Vert {#2} \right\Vert^{#1}}
\newcommand{\ind}[1]{{\mathbbm{1}{\left\{#1\right\}}}}

\newcommand{\s}{\mathrm{s}}

\renewcommand{\a}{\mathrm{a}}

\renewcommand{\d}{\mathrm{d}}
\renewcommand{\b}{\mathrm{b}}

\newcommand{\e}{\mathrm{e}}

\newcommand{\ic}{\mathrm{ic}}

\newcommand{\amd}{\mathrm{amd}}
\newcommand{\afp}{\mathrm{afp}}
\newcommand{\smd}{\mathrm{smd}}
\newcommand{\sfp}{\mathrm{sfp}}

\renewcommand{\vec}{\bm}

\newcommand{\EbNo}{E_\b/N_0}

\newcommand{\revise}[1]{{#1}} 
\newcommand{\revisee}[1]{{#1}}

\DeclareMathOperator*{\argmax}{arg\,max}
\DeclareMathOperator*{\argmin}{arg\,min}
\DeclareMathOperator*{\minimize}{minimize}

\newcommand{\T}{{\scriptscriptstyle\mathsf{T}}}

\newcommand{\Ac}{{\mathcal A}}
\newcommand{\Bc}{{\mathcal B}}
\newcommand{\Cc}{{\mathcal C}}

\newcommand{\Mc}{{\mathcal M}}
\newcommand{\Nc}{{\mathcal N}}

\newcommand{\Tc}{{\mathcal T}}
\newcommand{\Uc}{{\mathcal U}}
\newcommand{\Wc}{{\mathcal W}}

\newcommand{\EE}{\mathbb{E}}

\newcommand{\PP}{\mathbb{P}}
\newcommand{\RR}{\mathbb{R}}

\newcommand{\ppc}[1]{\textcolor{black}{#1}}
\newcommand{\bsc}[1]{\textcolor{black}{#1}}
\newtheorem{defi}{Definition}
\newtheorem{theo}{Theorem}

\newtheorem{lemma}{Lemma}
\newtheorem{remark}{Remark}

\newacronym{FA}{FA}{false alarm}
\newacronym{MD}{MD}{misdetection}
\newacronym{FP}{FP}{false positive}
\newacronym{aMD}{AMD}{alarm misdetection}
\newacronym{aFP}{AFP}{alarm false positive}
\newacronym{sMD}{SMD}{standard misdetection}
\newacronym{sFP}{SFP}{standard false positive}
\newacronym{MAC}{MAC}{multiple access channel}
\newacronym{UMA}{UMA}{unsourced multiple access}
\newacronym{HOMA}{H-OMA}{heterogeneous orthogonal multiple access}
\newacronym{HNOMA}{H-NOMA}{heterogeneous nonorthogonal multiple access}
\newacronym{AWGN}{AWGN}{additive white Gaussian noise}
\newacronym{SNR}{SNR}{signal-to-noise ratio}
\newacronym{wlog}{w.l.o.g.}{without loss of generality}
\newacronym{GMI}{GMI}{generalized mutual information}
\newacronym{RAN}{RAN}{radio access network}
\newacronym{RCUs}{RCUs}{random-coding union bound with parameter $s$}
\newacronym{IoT}{IoT}{Internet of Things}
\newacronym{ML}{ML}{maximum likelihood}
\newacronym{SIC}{SIC}{successive interference cancellation}
\newacronym{mMTC}{mMTC}{massive machine-type communications}
\newacronym{URLLC}{URLLC}{ultra-reliable low-latency communication}
\newacronym{eMBB}{eMBB}{enhanced mobile broadband}

\IEEEoverridecommandlockouts
\begin{document}
	\title{Unsourced Multiple Access With Common Alarm Messages: Network Slicing for Massive and Critical IoT}
	
	\author{
		\IEEEauthorblockN{Khac-Hoang Ngo, \emph{Member, IEEE,} Giuseppe Durisi, \emph{Senior Member, IEEE,} \\ Alexandre Graell i Amat, \emph{Senior Member, IEEE,}
			Petar~Popovski, \emph{Fellow, IEEE,} \\ Anders~E.~Kal{\o}r, \emph{Member, IEEE,} and Beatriz~Soret, \emph{Senior Member, IEEE} 
   }
		\thanks{Khac-Hoang Ngo, Giuseppe Durisi, and Alexandre Graell i Amat are with the Department of Electrical Engineering, Chalmers University of Technology, 41296 Gothenburg, Sweden (e-mails: {\{ngok, durisi, alexandre.graell\}@chalmers.se}). Petar~Popovski and Anders~E.~Kal{\o}r are with the Department of Electronic Systems, Aalborg University, 9220 Aalborg Øst, Denmark (e-mails: { \{petarp, aek\}@es.aau.dk}). Beatriz~Soret  is with the Telecommunication Research Institute (TELMA), Universidad de M\'alaga, 29010 M\'alaga, Spain (e-mail: {bsoret@ic.uma.es}).}
		\thanks{The work of Khac-Hoang Ngo was supported by the European Union's Horizon 2020 research and innovation programme under the Marie Sklodowska-Curie grant agreement No 101022113. The work of Giuseppe Durisi was partly supported by the Swedish Research Council under grant 2021-04970. The work of P. Popovski was supported by the Villum Investigator Grant “WATER” from the Velux Foundation, Denmark.
The work of A. E. Kalør was supported by the Independent Research Fund Denmark (IRFD) under Grant 1056-00006B.}
        \thanks{This paper was presented in part at the 56th Asilomar Conference on Signals, Systems, and Computers, Pacific Grove, CA, USA, 2022~\cite{Hoang2022hetorogeneous}.}
	}
	
	\maketitle
	
	\begin{abstract}
		We investigate the coexistence of massive and critical Internet of Things (IoT) services in the context of the  unsourced multiple access~(UMA) framework introduced by Polyanskiy (2017), where 
        all users employ 
        a common codebook and the receiver returns an unordered list of decoded codewords. \ppc{This setup is suitably modified to introduce heterogeneous traffic. Specifically, to model the massive IoT service, \revise{we assume that} a \emph{standard} message originates independently from each IoT device as in the standard UMA setup. To model the critical IoT service, we assume the generation of alarm messages that are \emph{common} for all devices. This setup requires a significant redefinition of the error events, 
        i.e., misdetections and false positives.} We further assume that the number of active users in each transmission attempt is random and unknown. We derive a random-coding achievability bound on the misdetection and false positive probabilities of both standard and alarm messages on the Gaussian multiple access channel. Using our bound, we demonstrate that orthogonal network slicing enables massive and critical IoT to coexist under the requirement of 
        high energy efficiency. On the contrary, we show that 
        nonorthogonal network slicing is energy inefficient due to the residual interference from the alarm signal when decoding the standard messages.
	\end{abstract}
	\begin{IEEEkeywords}
		Internet of things, unsourced multiple access, network slicing, random-coding bound, 
        misdetection, false positive
	\end{IEEEkeywords}
	
	\section{Introduction}\label{chap:introduction}
	The number of connected devices has grown drastically; it reached 13.2 billion in 2022, and is projected at 34.7 billion in 2028~\cite{Ericsson2022}. The data exchange between these devices gives rise to the \gls{IoT}~\cite{Colakovic2018}. 
	Two of the main segments of the \gls{IoT} landscape are massive \gls{IoT} and critical \gls{IoT}~\cite{Ericsson2020_cellular_IoT}. Massive \gls{IoT} connectivity targets a large number of low-cost, battery-limited, narrowband devices\textemdash meters, sensors, trackers, wearables\textemdash that transmit small data volumes in a sporadic and uncoordinated manner. 
    Critical \gls{IoT} connectivity aims to deliver data under strict latency and reliability guarantees for applications such as autonomous vehicles, real-time fault prevention, and real-time human-machine interaction. 	In a typical scenario in critical \gls{IoT}, multiple devices report a common malfunction or abnormal physical phenomenon, such as a gas leak or an out-of-range temperature, to an \gls{IoT} gateway. In the fifth-generation~(5G) wireless cellular standard, massive \gls{IoT} and critical \gls{IoT} are mapped to two separate use cases, named \gls{mMTC} and \gls{URLLC}, respectively~\cite{Chettri2020_5GIoT}. This paper aims to investigate the coexistence of massive and critical \gls{IoT} via an information-theoretic analysis.
	
	\subsubsection{State of the Art}
	Some of the key features of massive \gls{IoT} connectivity are captured by the recently proposed \gls{UMA} model~\cite{Polyanskiy2017}. This model differs from the classical multiple access setting in three fundamental aspects: i) all users transmit their messages using the same codebook and the decoder returns an unordered list of messages; ii) the error event is defined on a per-user basis as the event that the message transmitted by a given user is not included in the list produced by the decoder, and the error probability is averaged over all users; iii) each user sends a fixed amount of information within a finite-length frame. \revise{UMA is driven by the emergence of massive IoT applications, characterized by the deployment of millions of identical low-cost devices with codebooks hardwired in production. The common-codebook assumption eliminates the need for a codebook-assignment phase, which is implicitly assumed in classical multiple-access analyses but becomes impractical in the massive IoT setup. In UMA, the receiver decodes the list of transmitted messages without prior knowledge of the identity of the active devices~\cite{Polyanskiy_NASIT2018,Wu2020_massiveAccess,Shao2020,Polyanskiy_ISIT2021_tutorial,Che2023UMA}.} Under the \gls{UMA} framework, traditional as well as modern random access protocols~\cite{Berioli2016NOW} provide achievability results. 
	In~\cite{Polyanskiy2017}, a random-coding bound on the energy efficiency achievable on the Gaussian \gls{MAC} was derived. Modern random access schemes exhibit a large gap to this bound. This triggered \revise{a line} of research aimed at devising new coding schemes approaching the
	bound. \revise{Recent schemes are based on, e.g., coded compressed sensing~\cite{Amalladinne2020,fengler2019sparcs,Amalladinne2022,Ebert2022codedDemixing}, a combination of coded slotted ALOHA and conventional channel codes~\cite{Marshakov2019,Vem2019}, random spreading~\cite{Pradhan2020}, sparse Kronecker product~\cite{Han2021}, and tensor decomposition~\cite{Decurninge2020}}. The \gls{UMA} framework has been extended to the quasi-static fading channel~\cite{Kowshik2020energy}, \revise{the frequency-selective fading channel~\cite{Ozates2023},} the multiple-antenna channel~\cite{Fengler2019nonBayesian,Shyianov2021}, \revise{and a setting with variable-length codes and feedback~\cite{Yavas2021}}. 
	An extension to the case of random and unknown number of active users was presented in~\cite{Ngo2021ISITmassive,Ngo2022}, where both \glspl{MD}, i.e., transmitted messages that are not included in the decoded list, and \glspl{FP},\footnote{In~\cite{Ngo2021ISITmassive,Ngo2022}, the event that a message has not been transmitted but is included in the list produced by the decoder is called a {\em false alarm}. Here, we use the term {\em false positive} to avoid confusion with the {\em alarm} event.}
	i.e., decoded messages that are not transmitted, were considered.
	
	Both massive and critical \gls{IoT} can be analyzed under the framework of finite-blocklength information theory~\cite{Polyanskiy2010,Durisi2016toward}. For massive IoT, this framework accounts for the fact that the users transmit over a finite-length frame a finite number of bits. For critical IoT, finite-blocklength information theory provides accurate upper and lower 
    bounds on the maximum information rate that can be achieved under a given latency and reliability requirements. 
	
	Different \gls{IoT} traffic types typically need to coexist~\cite{Ericsson2020_cellular_IoT}. 
	In~\cite{Popovski2018}, the authors proposed to leverage reliability diversity to perform simultaneous transmission of different traffic types (also referred to as nonorthogonal network slicing) followed by successive interference cancellation. They showed that this approach leads to significant gains over orthogonal slicing when  \gls{mMTC} and \gls{eMBB} traffic are present, or when \gls{URLLC} and \gls{eMBB} traffic are present. However, they noted that nonorthogonal network slicing between \gls{URLLC} and \gls{mMTC} may be problematic due to the need to ensure reliability for \gls{URLLC} devices in the presence of the random interference patterns caused by \gls{mMTC} transmissions. 
 
    \bsc{The evolution toward more complex \gls{IoT} devices results in scenarios where each user generates heterogeneous traffic that can be critical or not critical.} \revise{In~\cite{Tominaga2023}, the authors investigated massive multiple-input multiple-output (MIMO) deployments for critical alarm traffic and noncritical \gls{mMTC} traffic, but did not consider the coexistence of both traffic types.}
	A first attempt to incorporate critical \gls{IoT} traffic into the \gls{UMA} model was undertaken in~\cite{Stern2019}. 
	There, on top of standard messages, the users communicate a common alarm message that needs to be decoded with higher reliability than the standard messages. 
	The authors assumed that a user drops the standard message in favor of the alarm message when both messages are present, and that the total number of active users transmitting either message is known. They showed that, in nonorthogonal network slicing, the \gls{FP} probability of alarm messages dominates and significantly reduces the energy efficiency  when the total number of users is large.
	
	\subsubsection{Contribution}
	In this paper, we \bsc{generalize} the \gls{UMA} setup with common alarm message proposed in~\cite{Stern2019} and study both orthogonal and nonorthogonal network slicing. In orthogonal slicing, standard and alarm messages are transmitted in separate blocks within a frame; in nonorthogonal slicing, both messages are transmitted over the whole frame. Differently from~\cite{Stern2019}, we consider a random and unknown number of active users for both traffic types, and that both messages are transmitted if they are present. \ppc{This means that, in our setup, the spectral efficiency is not automatically decreased upon the occurrence of alarm message, unlike~\cite{Stern2019}, where the suppression of the standard messages directly decreases the nominal rate of information conveyed through the system.} For both orthogonal and nonorthogonal network slicing, we provide a random-coding bound on the \gls{MD} and \gls{FP} probabilities of standard and alarm traffic, achievable on the Gaussian \gls{MAC}. 
    \revise{Note that for the alarm traffic, we need to use a different bounding technique compared to~\cite{Polyanskiy2017,Ngo2022} because the assumption of a common message transmitted by all active users is not compatible with the assumption of independent message generation used in~\cite{Polyanskiy2017,Ngo2022}.}
	We use our bounds to evaluate the achievable energy efficiency, measured by the minimum average energy per bit ($\EbNo$)
	required to satisfy given requirements on the \gls{MD} and \gls{FP} probabilities. Specifically, we let the standard traffic operate at a larger $\EbNo$ 
    than the minimum $\EbNo$ required when the alarm traffic is not present. We refer to the additional $\EbNo$ as backoff. We then report the minimum required $\EbNo$ for the alarm traffic. 
	
	For orthogonal network slicing, we investigate the impact of the probability that a user detects the alarm, which models the user sensitivity and limits the probability that the user transmits the alarm message when an alarm is present. We also study the impact of a constraint on the difference (in dB) between the power at which the standard and alarm codewords are transmitted, which we call the dynamic range. Through numerical results, we show that in orthogonal network slicing, a limited backoff is sufficient to transmit the alarm traffic with high energy efficiency, provided that i) the users are highly sensitive to the alarm, i.e., a large number of users detect and transmit the alarm message and ii) the dynamic range is large, i.e., the power at which the alarm message is transmitted is much smaller than that at which the standard message is transmitted. 
	We also show that nonorthogonal network slicing is inefficient because of the residual interference from the alarm message when decoding the standard messages. Specifically, for a small $\EbNo$ backoff of the standard message, nonorthogonal network slicing cannot satisfy the reliability requirements of both traffic types unless the  number of users transmitting the alarm message is reliably estimated, which occurs if all users transmit the alarm message or the transmit power of the alarm codeword is comparable to that of the standard codeword. In both cases, however, the required alarm $\EbNo$ is significantly higher than that of orthogonal network slicing. This confirms that reliability diversity~\cite{Popovski2018} between critical and massive \gls{IoT} is hard to exploit.  

	\subsubsection{Paper Organization}  
	The remainder of the paper is organized as follows. In Section~\ref{sec:systemModel}, we present the system model for \gls{UMA} with common alarm messages and define a random-access code. In Sections~\ref{sec:HOMA} and~\ref{sec:HNOMA}, we provide a random-coding bound for orthogonal and nonorthogonal network slicing, respectively. In Section~\ref{sec:numerical}, we present numerical results and discussions. We conclude the paper and provide some directions for future work in Section~\ref{sec:conclusion}. The proofs of our bounds can be found in the appendices. 
	
	\subsubsection{Notation}
	We denote system parameters by sans-serif letters, such as~$\sf{K}$, scalar random variables by upper case letters, such as~$X$, and their realizations by lower case letters, such as~$x$.  Vectors are denoted likewise with boldface letters, e.g., a random vector $\vec X$ and its realization~$\vec x$.   
	We denote the $n\times n$ identity matrix by $\vec I_n$, and the all-zero vector by $\mathbf{0}$. 
	The Euclidean norm and the transpose of $\vec x$ are 
    $\|\vec x\|$ and $\vec x^\T$, respectively. 
	Calligraphic uppercase letters, such as~$\Ac$, denote sets or events. We use $|\Ac|$ to denote the cardinality of $\Ac$ and $\mathfrak{P}(\Ac)$ the set of all subsets of $\Ac$, 
	$[m:n] \triangleq \{m,m+1,\dots,n\}$, 
	$[n] \triangleq [1:n]$, 
	$\ind{\cdot}$ is the indicator function, $\bar{\Ac}$ is the complement of the event $\Ac$, and $\RR$ denotes the set of real numbers. 
	We denote the Gamma function by $\Gamma(x) \triangleq \int_{0}^{\infty}z^{x-1}e^{-z}{\rm d}z$, and the upper incomplete Gamma functions by 
	$\Gamma(x,y) \triangleq \int_{y}^{\infty}z^{x-1}e^{-z} {\rm d} z$. We denote the Binomial distribution with parameters $(n,p)$ by $\mathrm{Bino}(n,p)$, and its probability mass function evaluated at $k$ by $\mathrm{Bino}(k;n,p) \triangleq \binom{n}{k}p^k(1-p)^{n-k}$. 
	Finally, $\Nc(\vec \mu,\vec \Sigma)$ denotes the multivariate real-valued Gaussian distribution with mean $\vec \mu$ and covariance matrix~$\vec \Sigma$. 
	
	
	\section{System Model}\label{sec:systemModel}
	We consider a \gls{MAC} in which  $\sf{K}$ users are given access opportunity over a frame consisting of $\sf{n}$ uses of a stationary memoryless \gls{AWGN} channel. \revisee{This channel model is relevant, e.g., in a time-division duplexing system where the base station broadcasts a downlink pilot signal, each user estimates its channel based on the pilot signal, and active users  pre-equalize their uplink signals based on the channel estimate~\cite{Mei2022,Qiao2023}. Here, as in~\cite{Qiao2023}, we assume that the channel estimation and pre-equalization steps are perfect and lead to a Gaussian channel with known \gls{SNR}, equal across all devices. We thus focus on the uplink transmission.}
	Let $\vec S_k \in \RR^\sf{n}$ be the signal transmitted by user~$k$, which is $\mathbf{0}$ if the user is inactive. This signal is subject to the power constraint $\|\vec S_k\|^2/\sf{n} \le \sf{P}$, $\forall k \in [{\sf{K}}]$. The  corresponding received signal is given by
	\begin{equation}
		\vec Y = \sum_{k = 1}^{\sf{K}} \vec S_k + \vec Z, \label{eq:model}
	\end{equation}
	where~$\vec Z \sim \Nc(\vec 0, \vec I_\sf{n})$ is the \gls{AWGN}, which is independent of $\{\vec S_k\}_{k=1}^{\sf{K}}$. \revise{Contrary to many \gls{UMA} studies, we assume that the number of transmitting users is random and unknown to the receiver.}

	\subsection{Message Generation} 
	\label{sec:msg_gen}
\begin{figure*}[tb]
		\centering

		\def\R{1.2}
		\scalebox{1}{\begin{tikzpicture}[font=\small]
				\draw[gray!50,dashed] (-9*\R,-1*\R) -- (5*\R,-1*\R);
				\draw[gray!50,dashed] (-9*\R,-2*\R) -- (5*\R,-2*\R);
				
				\node[right,text width=4cm] at (-9*\R,-.5*\R) () {Alarm event ($W_\a$)\\ occurred?};
				\node[right,text width=4cm] at (-9*\R,-1.5*\R) () {Standard message \\ generated?};
				\node[right,text width=4cm] at (-9*\R,-2.5*\R) () {Alarm message \\ generated?};
				
				\draw[-latex,ultra thick] (0-.1*\R,0) -- node[pos=.5,left=.4cm] () {Yes} (-3*\R,-1*\R); 
				\draw[-latex,ultra thick] (0-.1*\R,0) -- node[pos=.5,right=.4cm] () {No} (3*\R,-1*\R); 
				\draw[-latex,ultra thick] (-3*\R,-1*\R) -- node[pos=.5,left=.2cm] () {Yes, $\rho_\s$} 
				(-5*\R,-2*\R);
				\draw[-latex,ultra thick] (-3*\R,-1*\R) -- node[pos=.5,right=.2cm] () {No, $1 - \rho_\s$} 
				(-1*\R,-2*\R);
				
				\node[below] at (-6*\R,-3*\R) (MsMa) {\large $W_{\a,k}  \,=\, W_\a \atop W_{\s,k} \,\in\, \Mc_\s$};
				\node[below] at (-4*\R,-3*\R) (Ms1) {\large $W_{\a,k} \,=\, w_\e \atop W_{\s,k} \,\in\, \Mc_\s$};
				\draw[-latex,ultra thick] (-5*\R,-2*\R) -- node[pos=.5,left] () {Yes, $\rho_\d$} 
				(MsMa.north);
				\draw[-latex,ultra thick] (-5*\R,-2*\R) -- node[pos=.5,right] () {No, $1 - \rho_\d$} 
				(Ms1.north);
				
				\node[below] at (-2*\R,-3*\R) (Ma) {\large $W_{\a,k}  \,=\, W_\a \atop W_{\s,k} \,=\, w_\e$};
				\node[below] at (0*\R,-3*\R) (empty1) {\large $W_{\a,k} \,=\, w_\e \atop W_{\s,k} \,=\, w_\e$};
				\draw[-latex,ultra thick] (-1*\R,-2*\R) -- node[pos=.5,left] () {Yes, $\rho_\d$} 
				(Ma.north);
				\draw[-latex,ultra thick] (-1*\R,-2*\R) -- node[pos=.5,right] () {No, $1 - \rho_\d$} 
				(empty1.north);
				
				\node[below] at (2*\R,-2*\R) (Ms2) {\large $W_{\a,k} \,=\, w_\e \atop W_{\s,k} \,\in\, \Mc_\s$};
				\node[below] at (4*\R,-2*\R) (empty2) {\large $W_{\a,k} \,=\,w_\e \atop W_{\s,k} \,=\, w_\e$};
				\draw[-latex,ultra thick] (3*\R,-1*\R) -- node[pos=.5,left=.1cm] () {Yes, $\rho_\s$}  
				(Ms2.north);
				\draw[-latex,ultra thick] (3*\R,-1*\R) -- node[pos=.5,right] () {No, $1 - \rho_\s$} 
				(empty2.north);
			\end{tikzpicture}
		}
		\caption{A tree representation of the message generation for user~$k$. 
			The user generates the alarm message $W_{\a,k}$ and the standard message $W_{\s,k}$ indicated by the leaves. When a message is not generated, we say that the user picks the null message $w_\e$.}
		\label{fig:ModelVisualization}
	\end{figure*}
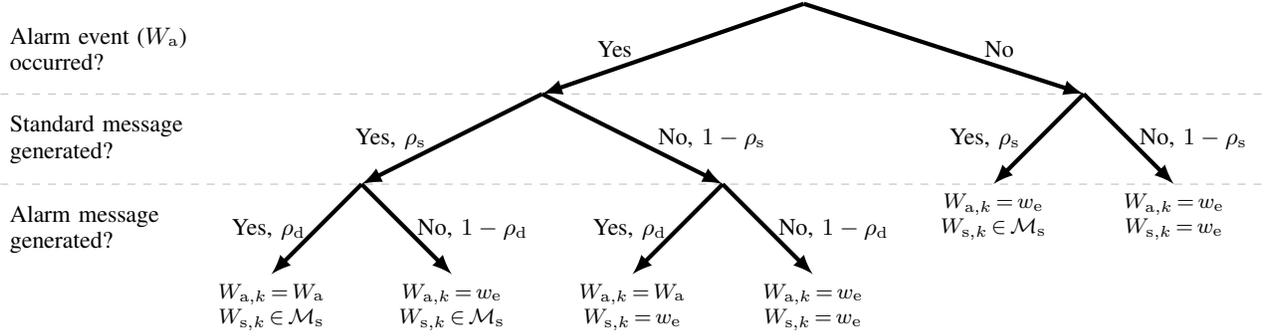

	We let $\Mc_\a$ denote the set of alarm messages and $\Mc_\s$ the set of standard messages; both sets are common to all users. Let $\sf{M}_\a \triangleq \vert\Mc_\a\vert$ and $\sf{M}_\s \triangleq \vert\Mc_\s\vert$. 
	In a frame, 
	if an alarm event has occurred, let $W_\a$ be the corresponding alarm message, drawn uniformly from $\Mc_\a$. Each user transmits this message with probability $\rho_\d$. 
	With probability $\rho_{\s}$, user $k\in [{\sf{K}}]$ generates a standard message $W_{\s,k}$ uniformly over $\Mc_\s$ and independently of the other users. To summarize, each user either transmits an alarm message, a standard message, both messages, or is inactive. Fig.~\ref{fig:ModelVisualization} illustrates the message generation rule. For convenience, we denote by $w_\e$ the ``null message'', mapped to the all-zero codeword (no transmission). 

    \begin{remark} \label{rem:rho_d_max}
        \revise{$\rho_\d$ can be expressed as
        $\rho_\d = \rho_{\d,\max} \rho_{\d,\text{trans}}$ where $\rho_{\d,\max}$ is the  probability that a user detects the alarm and $\rho_{\d,\text{trans}}$ is the probability that, upon detecting the alarm, the user decides to transmit the alarm message. 
		The probability $\rho_{\d,\max}$ models the user sensitivity to the alarm, while $\rho_{\d,\text{trans}}$ is a design parameter. Therefore, $\rho_\d$ is upper-bounded by~$\rho_{\d,\max}$.} 
    \end{remark}
	\begin{remark} \label{rem:std_vs_alarm}
		We assume that $\sf{M}_\s$ is much larger than $\sf{M}_\a$. In typical IoT scenarios, $\sf{M}_\s$ is in the order of $2^{100}$ (see, e.g., ~\cite{Polyanskiy2017} and~\cite[Rem.~3]{Zadik2019}), whereas $\sf{M}_\a$ can be less than~$10$, i.e., only a few different alarm events can occur. Since reporting the alarm is crucial for the system operation, the alarm message needs to be decoded with much higher reliability than the standard messages.
	\end{remark}
	
	Note that the number of users generating an alarm message and/or a standard message is random. 
	We assume that this number and the identity of the active users are unknown to the receiver. Hereafter, a user generating an alarm message is called an alarm user, and a user generating a standard message is called a standard user. A user can be simultaneously an alarm user and a standard user.
	
	\subsection{Random-Access Code}
	For the standard traffic, similar to~\cite{Polyanskiy2017}, all users employ the same codebook and the receiver decodes up to a permutation of the messages. Furthermore, as in~\cite{Ngo2022}, to address a random and unknown number of active users, we account for both \gls{MD} and \gls{FP} of the standard messages, referred to as SMD and SFP, respectively. \revise{An {SMD} occurs if a transmitted standard message is not included in the list of decoded standard messages. An {SFP} occurs if a message in the list of decoded standard messages has not been transmitted.} 
    We also consider \gls{MD} and \gls{FP} of the alarm message, referred to as AMD and AFP, respectively. \revise{An {AMD} occurs if an alarm event takes place but the receiver decodes the alarm message erroneously. An {AFP} occurs if the receiver returns an alarm message while no alarm event occurs.} We define the probabilities of these events and the random-access code below.
	
	\begin{defi}[Random-access code] \label{def:code}
		Consider the Gaussian \gls{MAC} with both standard and alarm traffic described above. An $(\sf{M}_\a, \sf{M}_\s, n, \epsilon_{\amd}, \epsilon_{\afp}, \epsilon_{\smd}, \epsilon_{\sfp})$ random-access code for this channel, where $\sf{M}_\a$ and $\sf{M}_\s$ are the sizes of the alarm and standard message sets, respectively, $\sf{n}$ is the framelength, and $\epsilon_{\amd}, \epsilon_{\afp}, \epsilon_{\smd}, \epsilon_{\sfp} \in (0,1)$, consists of: 
		\begin{itemize}
			\item A random variable $U$ defined on a set $\Uc$ that is revealed to both the users and the receiver before the transmission. This random variable acts as common randomness and allows for the use of randomized coding strategies.
			
			\item An encoding function $$f \colon \Uc \times (\Mc_\a \cup \{w_\e\}) \times (\Mc_\s\cup \{w_\e\}) \to \RR^n$$ that produces the transmitted codeword $\vec S_k = f(U, W_{\a,k}, W_{\s,k})$ for user~$k$,  for a given alarm message $W_{\a,k}$ and standard message $W_{\s,k}$. 
			
			\item A decoding function $$g \colon \Uc \times \RR^n \to (\Mc_\a \cup \{w_\e\}) \times (\mathfrak{P}(\Mc_\s) \cup \{w_\e\})$$ that provides an estimate $\widehat{W}_{\a}$ of the common alarm message $W_\a$ and an estimate $\widehat{\Wc}_\s = \{\widehat{W}_{\s,1}, \dots, \widehat{W}_{\s,|\widehat{\Wc}|}\}$ of the list of transmitted standard messages. That is, $(\widehat{W}_{\a}, \widehat{\Wc}_\s) = g(U,\vec Y)$.
		\end{itemize}
		Let $\widetilde{\Wc}_\s = \{\widetilde{W}_{\s,1},\dots,\widetilde{W}_{\s,|\widetilde{\Wc}|}\}$ be the set of distinct elements of $\Wc_\s = \{W_{\s,k} \colon W_{\s,k} \ne w_\e, k \in [{\sf{K}}]\}$. 
		Let $\Ac$ denote the event that an alarm occurs. We assume that the decoding function satisfies the following constraints on the AMD, AFP, SMD, and SFP probabilities, respectively:
		\begin{align}
			P_\amd \triangleq \P{ \widehat{W}_{\a} \ne W_\a \big\vert \Ac} &\leq \epsilon_\amd, \label{eq:cond_aMD}\\
			P_\afp \triangleq \P{ \widehat{W}_{\a} \ne w_\e \big\vert \bar{\Ac}} &\leq \epsilon_\afp,
			\label{eq:cond_aFP} \\
			P_{\smd \vert \Bc} \triangleq \EE_{|\widetilde{W}_\s|}\Bigg[\frac{1}{|\widetilde{\Wc}_\s|}\sum_{i=1}^{|\widetilde{W}_\s|}\P{\widetilde{W}_{\s,i}\notin \widehat{\mathcal{W}}_\s \big\vert \Bc}\Bigg] &\leq \epsilon_\smd, 
            \label{eq:cond_sMD} \\
			P_{\sfp \vert \Bc} \triangleq \EE_{|\widehat{\mathcal{W}}_\s|}\Bigg[\frac{1}{|\widehat{\mathcal{\Wc}}_\s|}\sum_{i=1}^{\widehat{\mathcal{W}}_\s}\P{\widehat{W}_{\s,i}\notin \widetilde{\Wc}_\s \big\vert \Bc}\Bigg] &\leq \epsilon_\sfp. 
            \label{eq:cond_sFP}
		\end{align}
		Here, we used the convention that $0/0 = 0$ to circumvent the cases $|\widetilde{\Wc}_\s| = 0$ or $|\widehat{\Wc}_\s| = 0$. Furthermore, 
		   \eqref{eq:cond_sMD} and \eqref{eq:cond_sFP} hold for both $\Bc = \Ac$ and $\Bc = \bar{\Ac}$.  
	\end{defi}
	
    
		Our definition of a random-access code differs from that in~\cite[Def.~2]{Stern2019} in two aspects. First, \cite[Def.~2]{Stern2019} applies exclusively to the case where a user drops the standard message in favor of the alarm message when both messages are present, and this does not result in a SMD. Here, we assume that both messages are transmitted if they are present. 
		Second, while~\cite[Def.~2]{Stern2019} assumes a known total number of active users and considers only MD for the standard traffic, we consider an unknown number of active users and account for both SMD and~SFP. 
	
	\begin{remark} \label{rem:rho_d}
		Note that $\rho_\d$ should  be sufficiently large to satisfy the reliability target of the alarm traffic. Specifically, $P_\amd$ is lower-bounded by the probability that no user transmits the alarm message, $(1-\rho_\d)^{\sf{K}}$. Thus, to guarantee $P_\amd \le \epsilon_\amd$, one must have that $(1-\rho_\d)^{\sf{K}} \le \epsilon_\amd$,  i.e., 
		$\rho_\d \ge 1 - \epsilon_{\amd}^{1/\sf{K}}$.
	\end{remark}
	
		In the next section, we shall use a random-coding argument to obtain achievability
		bounds, i.e., upper bounds on the error probabilities in~\eqref{eq:cond_aMD}--\eqref{eq:cond_sFP}. Specifically, we will construct a codebook ensemble for which~\eqref{eq:cond_aMD}--\eqref{eq:cond_sFP} hold in
		average. Unfortunately, this does not imply that there exists a code in this ensemble that satisfies all these four constraints simultaneously. The introduction of the random variable~$U$ in Definition~\ref{def:code} allows us to circumvent this issue by enabling randomized coding strategies. Specifically, by proceeding as
		in~\cite[Th. 19]{Polyanskiy2011feedback}, one can show that there exists a randomized coding strategy that achieves~\eqref{eq:cond_aMD}--\eqref{eq:cond_sFP} simultaneously and involves time-sharing among at most five deterministic codes (i.e., $|\Uc| \le 5$) in
		this ensemble.
	
	
	\section{Heterogeneous Orthogonal Multiple Access}\label{sec:HOMA}
\begin{figure}[t!]
        \centering
        \subfigure[\revise{In \gls{HOMA}, a frame comprises two blocks: an alarm block containing $\sf{n}_\a$ channel uses dedicated to the alarm traffic, and a standard block containing $\sf{n}_\s$ channel uses dedicated to the standard traffic. Note that $\sf{n}_\s + \sf{n}_\a$ is not necessarily equal to $\sf{n}$, i.e., we allow for empty slots.}]{
            \scalebox{.76}{\begin{tikzpicture} 
                \draw[step=.5cm,black,thin] (0,0) grid (10,.5);
                \draw[latex-latex] (0,.75) -- node[above,midway] () {$\sf{n}$} (10,.75);
                
                \draw[step=.5cm,draw=black,pattern={north east lines},pattern color=red] (0,0) grid (2.5,.5) rectangle (0,0);
                
                \draw[step=.5cm,draw=black,pattern={north west lines},pattern color=blue] (2.5,0) grid (9.5,.5) rectangle (2.5,0);
    
                \draw[step=.5cm,draw=black,fill=white] (9.5,0) grid (10,.5) rectangle (9.5,0);
    					
                \draw[latex-latex] (0,-.25) -- node[below,midway, align=center] () {{$\sf{n}_\a$} \\ alarm block} (2.5,-.25);
                
    			\draw[latex-latex] (2.5,-.25) -- node[below,midway, align=center] () {{$\sf{n}_\s$} \\ standard block} (9.5,-.25);
    		\end{tikzpicture}}
            \label{fig:HOMA}
        }
        \hspace{.2cm}
        \subfigure[\revise{In \gls{HNOMA}, a superposition of both the standard and alarm codewords is transmitted over the whole frame.}]{
            \scalebox{.76}{\begin{tikzpicture} 
                \draw[step=.5cm,black,thin] (0,0) grid (10,.5);
                \draw[latex-latex] (0,.75) -- node[above,midway] () {$\sf{n}$} (10,.75);
                
                \draw[step=.5cm,draw=black,pattern={north east lines},pattern color=red] (0,0) grid (10,.5) rectangle (0,0);
                \draw[step=.5cm,draw=black,pattern={north west lines},pattern color=blue] (0,0) grid (10,.5) rectangle (0,0);
            \end{tikzpicture}
            \label{fig:HNOMA}}
        }
        \caption{\revise{Illustration of a frame in the two considered network slicing strategies, namely, \gls{HOMA} and \gls{HNOMA}.}}
    \end{figure}
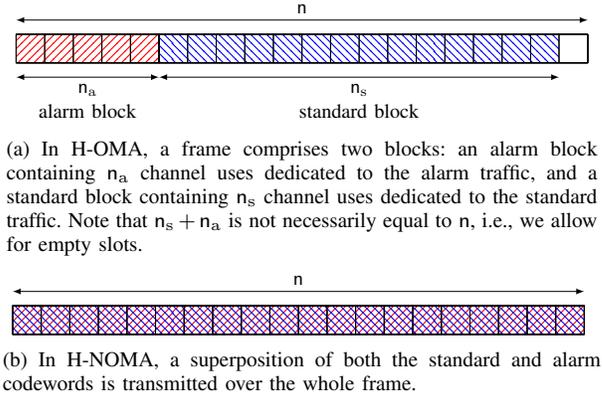

	We analyze next an orthogonal network slicing strategy, which we refer to as~\gls{HOMA}. Each frame \revise{comprises} two blocks: one containing $\sf{n}_\a$ channel uses dedicated to the alarm traffic, and the other containing $\sf{n}_\s~\revise{\le}~\sf{n}-\sf{n}_{\mathrm{a}}$ channel uses dedicated to the standard traffic. \revise{We illustrate the frame structure in Fig.~\ref{fig:HOMA}.} We next describe the signal model in each block. 
	
	\subsection{Signal Model}
    We assume that the users share an alarm codebook containing $\sf{M}_\a$ codewords of length $\sf{n}_\a$ and a standard codebook containing $\sf{M}_\s$ codewords of length $\sf{n}_\s$.
    
	\subsubsection{Alarm Block} \label{sec:h-omaAlarmModel}
	If an alarm event has occurred, the common alarm message $W_\a$ is sent in the alarm block by every user that detects the alarm and decides to transmit (i.e., with probability $\rho_\d$). 
    Let $\vec X_\a \in \RR^{\sf{n}_\a}$ be the alarm codeword corresponding to $W_\a$.
	The received signal is 
		\begin{equation}
		\vec Y_{\mathrm{a}} = K_{\mathrm{a}}\vec X_\a + \vec Z_{\mathrm{a}},
				\label{eq:slicedSignalAlarm}
			\end{equation}
	where $K_{\a} \ge 0$ is the number of alarm users 
	and $\vec Z_\a \sim\mathcal{N}(\vec 0,\vec I_{\sf{n}_\a})$ is the \gls{AWGN}. If no alarm event occurs, $K_\a = 0$; otherwise, $K_\a \sim {\rm Bino}(\sf{K},\rho_\d)$. We impose the power constraint $\norm{\vec X_\a}/\sf{n}_{\a} \leq \sf{P}_\a$. This model is equivalent to a single-user \gls{AWGN} channel with random \gls{SNR} $K_{\a}^2\sf{P}_\a$. 
	The average energy per bit of the alarm traffic is upper-bounded by $(\EbNo)_{\a} = 
	\frac{\sf{n}_{\a} \sf{P}_\a \rho_\d \sf{K}}{2\log_{2}\sf{M}_\a}$. 
	
	\subsubsection{Standard Block} 
	The standard block resembles the UMA channel with random and unknown number of active users considered in~\cite{Ngo2022}. 
	The number of active users in this block is $K_\s \sim {\rm Bino}(\sf{K},\rho_\s)$. Without loss of generality, we assume that the first $K_\s$ users are active. Let $\vec X_{\s,k} \in \RR^{\sf{n}_\s}$ be the standard codeword mapped from $W_{\s,k}$. The received signal is 
		\begin{equation}
		\vec Y_{\mathrm{s}} = \sum_{k=1}^{K_{\s}}\vec X_{\s,k} + \vec Z_{\mathrm{s}},
			\label{eq:slicedSignalStandard}
	       \end{equation}
	where 
    $\vec Z_{\mathrm{s}}\sim\mathcal{N}(\vec 0,\vec I_{\sf{n}_\s})$ is the \gls{AWGN}. We impose a power constraint $\norm{\vec X_{\s,k}}/\sf{n}_\s \leq \sf{P}_\s$, $k \in [K_{\s}]$. 
	The average energy per bit of the standard traffic is upper-bounded by $(\EbNo)_\s = 
	\frac{\sf{n}_\s\sf{P}_\s}{2\log_2\sf{M}_\s}$. 
	
	\subsubsection{Encoder, Decoder, and Power Constraint}
	In accordance with Definition~\ref{def:code}, the output of the encoding function is the concatenation of an alarm codeword and a standard codeword. The received signal $\vec Y$ over the frame is $\vec Y = [\vec Y_{\a}^\T \ \vec Y_{\s}^\T]^\T$. The decoder outputs an alarm message (or $w_\e$) and a list of standard messages (or $w_\e$) that are returned by the decoders operating on the two blocks, respectively. To satisfy the power constraint, we set $(\sf{P}_\a, \sf{P}_\s)$ such that $\sf{n}_\a \sf{P}_\a + \sf{n}_\s \sf{P}_\s \le \sf{n} \sf{P}$.
	We note that, for fixed $\sf{P}_\a$ and $\sf{P}_\s$, orthogonality implies that the \gls{MD} and \gls{FP} in the standard traffic are independent of the alarm event, i.e., $P_{\smd | \bar{\Ac}} = P_{\smd \vert \Ac}$ and $P_{\sfp | \bar{\Ac}} = P_{\sfp \vert \Ac}$.
\subsection{Random-Coding Bound} \label{sec:HOMA_scheme}
In the following, for a given frame split $(\sf{n}_\a, \sf{n}_\s)$ and power allocation $(\sf{P}_\a,\sf{P}_\s)$, we derive a random-coding bound on the AMD, AFP, SMD, and SFP probabilities \revise{defined} in \eqref{eq:cond_aMD}--\eqref{eq:cond_sFP}. 

\subsubsection{Alarm Block}
We fix \revise{a transmit power} $\sf{P}_\a' \le \sf{P}_\a$ and draw $\sf{M}_\a$ alarm codewords $\Cc_\a = \{\vec C_{\a,1},\ldots, \vec C_{\a,\sf{M}_\a}\}$ independently from $\mathcal{N}(\vec 0, \sf{P}'_\a\vec I_{\sf{n}_{\a}})$. To convey an alarm message $W_\a$, the alarm users transmit $\vec C_{\a, W_\a}$ provided that $\|\vec C_{\a,W_\a}\|^2 \le \sf{n}_{\a}\sf{P}_\a$. Otherwise, they transmit the all-zero codeword.\footnote{This implies that, if there is an alarm, no user transmits the alarm codeword if it violates the power constraint. In our bound, we use a change of measure to account for the probability that a codeword violates the power constraint. We choose $\sf{P}'_\a$ such that  this probability is small.} To summarize, $\vec X_\a = \vec C_{\a,W_\a} \ind{\|\vec C_{\a,W_\a}\|^2 \le \sf{n}_{\a}\sf{P}_\a}$.
Given a realization  $\vec y_{\a}$ of the received signal, the decoder proceeds in two steps. First, it detects if an alarm is present. \revise{Specifically,} it finds an initial estimate of the number of active alarm users as
\begin{equation} \label{eq:Ka_estimation}
	{K}_{\a}' = \argmax_{k \in \{0\} \cup [k_{\a,\ell}:k_{\a,u}]} m_\a(\vec y_{\a},k),
\end{equation}
where $m_\a(\vec y_\a,k) \triangleq  P_{\vec Y_\a \vert K_\a}(\vec y_\a \vert k) = (2\pi(1+k^2\sf{P}_\a))^{-\sf{n}_\a/2} \exp\big(-\frac{\|\vec y_\a\|^2}{2(1+k^2 \sf{P}_\a)}\big)$. The limits $k_{\a,\ell}$ and $k_{\a,u}$ are chosen such that $\sum_{k = k_{\a,\ell}}^{k_{\a,u}} {\rm Bino}(k;\sf{K},\rho_\d)$ exceeds a threshold, as discussed in Remarks~\ref{rem:alarm_decoding} and~\ref{rem:neyman_pearson} below. 
If $K_{\a}'=0$, the decoder declares that there is no alarm, i.e., it returns the null message $w_\e$. Otherwise, the decoder proceeds to decode the alarm message as
\begin{equation}
	(\widehat{W}_{\a}, \widehat{K}_\a) = \argmin_{w\in \Mc_\a,\ k \in \{0\} \cup [k_{\a,\ell}:k_{\a,u}]} \norm{\vec y_{\a} - k \revise{\vec C_{\a,w}}} \label{eq:alarm_decoding}
\end{equation}
Finally, the decoder returns $\widehat{W}_{\a}$ if $\widehat{K}_\a > 0$, or returns $w_\e$ if $\widehat{K}_\a = 0$.
\begin{remark} \label{rem:alarm_decoding}
	In~\cite{Stern2019}, the decoding of the alarm message consists only of a step similar to~\eqref{eq:alarm_decoding}, where the search space for $\widehat{K}_\a$ is $[0:\sf{K}]$. 
	If there is no alarm, then $K_\a = 0$, $\vec Y_\a \sim \Nc(\mathbf{0}, \vec I_{\sf{n}_\a})$, and $\vec Y_\a - \revise{\vec C_{\a,w}} \sim \Nc(\mathbf{0}, (1+\sf{P}'_\a)\vec I_{\sf{n}_\a})$. In this case, if $\sf{P}'_\a \ll 1$, i.e., $1+\sf{P}'_\a \!\approx\! 1$, then~\eqref{eq:alarm_decoding} outputs $\widehat{K}_\a = 1$ with significant probability. This causes a high AFP probability, which was the bottleneck in~\cite{Stern2019}. We overcome this bottleneck by adding the alarm-detection step~\eqref{eq:Ka_estimation}, 
	which aims to detect if $K_\a = 0$, i.e., no alarm \revise{is present}, or $K_\a \in [k_{\a,\ell}:k_{\a,u}]$, i.e., an alarm is present. We also include $k = 0$ in the refined estimation of $K_\a$ in~\eqref{eq:alarm_decoding} to be able to detect the no-alarm state in this step. 
\end{remark}
\begin{remark} \label{rem:neyman_pearson}
	Our alarm-detection step \revise{is} a relaxed version of the Neyman-Pearson binary hypothesis test between \revise{the following two hypotheses:} i) $\vec y_\a$ is generated from $\Nc(\mathbf{0},\vec I_{\sf{n}_\a})$ (no alarm), and ii) $\vec y_\a$ is generated from $\Nc(\mathbf{0},\vec (1+K_\a^2 \sf{P}_\a)\vec I_{\sf{n}_\a})$ with $K_\a \sim {\rm Bino}(\sf{K},\rho_\d)$ (an alarm is present). 
	This Neyman-Pearson test declares that there is no alarm if $$\Omega(\vec y_\a) \triangleq \frac{m_\a(\vec y_\a,0)}{\sum_{k = 0}^{\sf{K}} {\rm Bino}(k; \sf{K},\rho_\d) {m_\a(\vec y_\a,K_\a)}} \ge \tau$$ and that there is an alarm otherwise, for a \revise{suitably chosen} threshold~$\tau$. If $\sum_{k = k_{\a,\ell}}^{k_{\a,u}} {\rm Bino}(k;\sf{K},\rho_\d)$ is close to $1$,  we have that $$\Omega(\vec y_\a) \approx
		\bar{\Omega}(\vec y_\a) \triangleq \frac{\sum_{k = k_{\a,\ell}}^{k_{\a,u}} {\rm Bino}(k; \sf{K},\rho_\d) m_\a(\vec y_\a,0)}{\sum_{k = k_{\a,\ell}}^{k_{\a,u}} {\rm Bino}(k; \sf{K},\rho_\d) m_\a(\vec y_\a,k)}.
	$$
	With $\tau = 1$, a sufficient condition for $\bar{\Omega}(\vec y_\a) \ge \tau$ is that $m_\a(\vec y_\a,0) \ge m_\a(\vec y_\a,k)$ for all $k \in [k_{\a,\ell}:k_{\a,u}]$, which we use in our scheme to signify that there is no alarm. Here, we tune $k_{\a,\ell}$ to control the AFP, and introduce $k_{\a,u}$ to avoid large sums (up to $\sf{K}$) in the random-coding bound. While our test \revise{is} suboptimal \revise{compared} to the  Neyman-Pearson test, it simplifies the derivation of the bound.
\end{remark}

An error analysis of this scheme leads to the following upper bounds on $P_\amd$ and $P_\sfp$. 
\begin{theo}[Random-coding bound for the alarm block]\label{th:alarm_block}
	Fix $r_\a$, $\sf{n}_{\a} \in [\sf{n}]$, $k_{\a, \ell} \in [1:\sf{K}]$, $k_{\a, u} \in [k_{\a, \ell} + 1:\sf{K}]$, and $\sf{P}_\a' < \sf{P}_\a$. The AMD and AFP probabilities achieved by the random-coding scheme just described are upper-bounded by $\epsilon_{\amd}$ and $\epsilon_{\afp}$, respectively, where 
	\begin{align}		
		\epsilon_\amd 
		&= \sum_{k_\a = k_{\a, \ell} }^{k_{\a, u}} \!\!P_{K_\a}(k_\a) \Bigg(\zeta(k_\a,0) + \min\Bigg\{1,\sum_{k'_\a = k_{\a,\ell}}^{k_{\a, u}} \!\!\zeta(k_{\a},{k}'_{\a}) \Bigg\} \notag\\
        &\qquad \cdot \sum_{\hat{k}_\a \in \{0\} \cup [k_{\a,\ell} :k_{\a,u}]} \gamma_\amd(k_\a,\hat k_\a) \Bigg)+ \nu_0, \label{eq:RCU_AMD}\\
		\epsilon_\afp 
		&=  \min\Bigg\{1,\sum_{k_\a' = k_{\a,\ell}}^{k_{\a,u}} \zeta(0,{k}_\a') \Bigg\} \sum_{\hat{k}_{\a} = k_{\a,\ell}}^{k_{\a,u}} \gamma_\afp(\hat{k}_\a) ,  \label{eq:RCU_AFP} 
	\end{align}
	with 
	$P_{K_\a}(k_\a) = {\rm Bino}(\sf{K},\rho_\d)$ and
	\begin{align}
		\nu_0 &\triangleq  \frac{\Gamma(\sf{n}_\a/2,\sf{n}_\a \sf{P}_\a/(2\sf{P}'_\a))}{\Gamma(\sf{n}_\a/2)} + 1 - \sum_{k = k_{\a, \ell}}^{k_{\a,u}} P_{K_\a}(k), \label{eq:p0}\\ 
		\gamma_\amd(k_\a,{\hat{k}_{\a}}) &\triangleq \min_{s>0} \PP\bigg[\sum_{i=1}^{\sf{n}_\a} \imath_s(\hat{k}_\a, X'_i; Y'_i) \le \ln\frac{\sf{M}_\a\!-\!1}{V} \bigg], \label{eq:gamma_md} \\ 
		\gamma_\afp(\hat{k}_\a) &\triangleq \min_{s  > 0} \EE\bigg[\frac{1}{\Gamma(\sf{n}_\a/2)} \notag \\			& ~\cdot
		\Gamma\bigg(\frac{\sf{n}_\a}{2}, \frac{1}{2\beta} \bigg(\frac{\sf{n}_{\a}}{2}\ln(1 \!+\! 2\hat{k}_{\a}^2\sf{P}'_\a s ) - \ln \frac{\sf{M}_\a}{V}\bigg) \bigg) \bigg], \label{eq:gamma_fp}\\ 
		\beta &\triangleq s - s(1+2\hat{k}_{\a}^2\sf{P}'_\a s)^{-1}, \label{eq:beta} \\
		\zeta(k_\a,{k}_{\a}') &\triangleq \min_{k \in \{0\} \cup [k_{\a,\ell}:k_{\a,u}],  \atop k \ne{k}_{\a}'} \P{m_\a(\vec Y'_{\a},{k}_{\a}') \!>\! m_\a(\vec Y'_{\a},k)}. \label{eq:zeta} 
	\end{align}
	In~\eqref{eq:zeta}, $\vec Y'_{\a} \sim \mathcal{N}(\mathbf{0}, (1+k_\a^2 \sf{P}'_\a)\vec I_{\sf{n}_{\a}})$. In~\eqref{eq:gamma_md} and~\eqref{eq:gamma_fp}, $V$ is uniformly distributed over $[0,1]$. In~\eqref{eq:gamma_md}, $[X'_1 \dots X'_{\sf{n}_{\a}}]^\T \sim \mathcal{N}(\mathbf{0},\sf{P}'_\a\vec I_{\sf{n}_{\a}})$; given $X'_i = x'_i$, we have that $Y'_i \sim \mathcal{N}({k}_{\a} x'_i,1)$;  $\imath_s(\hat{k}_\a, X'_i;Y'_i)$ is the generalized information density given by
	\begin{equation} \label{eq:def_gen_info_den}
		\imath_s(\hat{k}_\a, x;y) \triangleq -s(y-{k}_{\a}x)^2 + \frac{sy^2}{1+2s\hat{k}_\a^2P'} + \frac{1}{2} \ln\big(1+2s\hat{k}_\a^2P'\big).
	\end{equation}
\end{theo}
\begin{IEEEproof} 
	\revise{The assumption that all alarm users transmit the same codeword is not compatible with the assumption of independent codeword generation in the original \gls{UMA} setup. Consequently, we need to use a different bounding technique than in~\cite{Polyanskiy2017,Ngo2022}. Specifically, the proof of Theorem~\ref{th:alarm_block} relies on the \gls{RCUs}~\cite{Martinez2011RCUs}.} See Appendix~\ref{proof:HOMA} for details.
\end{IEEEproof}

		The quantity $\zeta(k_{\a},{k}'_{\a})$ is an upper bound on the probability that the maximum-likelihood estimation step~\eqref{eq:Ka_estimation} returns~${k}'_{\a}$, given that there are $K_{\a} = k_\a$ alarm users. Closed-form expressions for $\zeta(k_{\a},{k}'_{\a})$ can be deduced from~\cite[Th.~2]{Ngo2022}.
	
		In the numerical experiments (Section~\ref{sec:numerical}), we use $s^* = \frac{1}{4} + \frac{\sqrt{\hat{k}_\a^4 (\sf{P}'_\a)^2 + 4{k}_\a^2 \sf{P}'_\a + 4} - 2}{4\hat{k}_\a^2 \sf{P}'_\a}$, which maximizes the generalized mutual information $\E{\sum_{i=1}^{\sf{n}_\a} \imath_s(\hat{k}_\a, X'_i; Y'_i) } =  \sf{n}_\a \Big[-s + \frac{s(1+{k}_\a^2 \sf{P}'_\a)}{1+2s\hat{k}_\a^2 \sf{P}'_\a} + \frac{1}{2} \ln\big(1+ 2s\hat{k}_\a^2 \sf{P}'_\a\big)\Big]$, as an initial value for~$s$ when performing the minimization in~\eqref{eq:gamma_md}. 
		We also use $1/4$, which minimizes an upper bound \revise{on} $\gamma_\afp(\hat k_\a)$, as an initial value for $s$ when performing the minimization in~\eqref{eq:gamma_fp}. 

\subsubsection{Standard Block}
We consider the random-coding scheme proposed in~\cite[Sec.~III-A]{Ngo2022}. 
Specifically, we fix \revise{a transmit power} $\sf{P}'_\s < \sf{P}_\s$ and generate the $\sf{M}_\s$ standard codewords $\Cc_\s = \{\vec C_{\s,1},\ldots, \vec C_{\s,\sf{M}_\s}\}$ independently from $\mathcal{N}(\vec 0, \sf{P}'_\s\vec I_{\sf{n}_\s})$. To convey a standard message $W_{\s,k}$, the $k$th standard user transmits  
$\vec X_{\s,k} = \vec C_{\s,W_{\s,k}} \ind{\|\vec C_{\s,W_{\s,k}}\|^2 \le \sf{n}_\s\sf{P}_\s}$. Given the channel output $\vec y_\s$, 
the decoder first estimates the number of active users as 

\begin{equation}
	{K}_{\s}' = \argmax_{k \in [k_{\s, \ell}:k_{\s, u}]} m_\s(\vec y_{\s},k) \,,\label{eq:Ks_estimation}
\end{equation}
where $m_\s(\vec y_{\s},k)$ is a suitably chosen metric;  
$k_{\s, \ell}$ and $k_{\s, u}$ are limits on $K'_\s$, chosen based on the distribution of $K_\s$, i.e., based on $\rho_\s$. Then, given $K'_\s \!=\! k'_\s$, the decoder chooses the output list as
\begin{equation}
		\widehat{\mathcal{W}} = \argmin_{\mathcal{W}'\subset [\Mc_\s] \colon |\mathcal{W}'| \in [\underline{k_\s'}:\overline{k_\s'}] } \| \vec y_{\s} - c(\Wc') \|^2, 
		\label{eq:std_decoding}
	\end{equation}
	where $\underline{k_\s'} \triangleq \max\{k_{\s,\ell},k_\s'-r_\s\}$, $\overline{k_\s'} \triangleq \min\{k_{\s,u},k_\s+r_\s\}$, $c(\mathcal{W}') \triangleq \sum_{i\in \mathcal{W}'} \revise{\vec C_{\s,i}}$, 
	and $r_\s$ is a chosen nonnegative integer, which we call the \emph{standard-message decoding radius}. 
	Bounds on the SMD and SFP probabilities achieved by this random-coding scheme follow from~\cite[Th.~1]{Ngo2022}.  We state these bounds in the next theorem.
	
	\begin{theo}[Random-coding bound for the standard block]  \label{th:std_block}
		Fix $r_\s$, $\sf{n}_\s \in [\sf{n}]$, $k_{\s, \ell} \in [0:\sf{K}]$, $k_{\s, u} \in [k_{\s, \ell}+1 :\sf{K}]$, and $\sf{P}_\s' < \sf{P}_\s$. The SMD and SFP probabilities achieved by the random-coding scheme just described are upper-bounded by $\epsilon_\smd$ and $\epsilon_{\sfp}$, respectively, where
		\begin{align}
			\epsilon_\smd &= 
			\sum_{k_\s = \max\{k_{\s,\ell},1\}}^{k_{\s,u}} P_{K_\s}(k_\s) \notag \\
        &\qquad\cdot \bar{\epsilon}_\smd(\sf{M}_\s,\sf{n}_\s,k_\s,\sf{P}_\s,\sf{P}'_\s,r_\s,k_{\s,\ell},k_{\s,u})   + \nu_1, \label{eq:RCU_OMA_SMD}\\
			\epsilon_\sfp &=  
			\sum_{k_\s =k_{\s,\ell}}^{k_{\s,u}} P_{K_\s}(k_\s) \bar{\epsilon}_\sfp(\sf{M}_\s,\sf{n}_\s,k_\s,\sf{P}_\s,\sf{P}'_\s,r_\s,k_{\s,\ell},k_{\s,u})  \notag \\
        &\qquad+ \nu_1, \label{eq:RCU_OMA_SFP}
		\end{align}	
		with 
		\begin{align}
			&\bar{\epsilon}_\smd(\sf{M}_\s,\sf{n}_\s,k_\s,\sf{P}_\s,\sf{P}'_\s,r_\s,k_{\s,\ell},k_{\s,u}) \notag \\
            &~ \triangleq \sum_{k_\s' = k_{\s,\ell}}^{k_{\s,u}} \sum_{t\in \Tc}\frac{t+(k_\s-\overline{k_\s'})^+}{k_\s} \min\{p_t,q_t, \xi(k_\s,k_\s')\}, \label{eq:bar_eps_SMD}\\
			&\bar{\epsilon}_\sfp(\sf{M}_\s,\sf{n}_\s,k_\s,\sf{P}_\s,\sf{P}'_\s,r_\s,k_{\s,\ell},k_{\s,u}) \notag \\
            &~\triangleq \sum_{k_\s' = k_{\s,\ell}}^{k_{\s,u}} \sum_{t\in \Tc} \sum_{t' \in \Tc_t} \frac{t'+(\underline{k_\s'}-k_\s)^+}{k_\s - t - {(k_\s - \overline{k_\s'})}^+ + t' + {(\underline{k_\s'}-k_\s)}^+} \notag \\
			&\qquad \qquad \qquad \quad \cdot \min\{p_{t,t'}, q_{t,t'}, \xi(k_\s,k_\s')\}. \label{eq:bar_eps_SFP}
		\end{align}
		Here, \revisee{in~\eqref{eq:RCU_OMA_SMD} and~\eqref{eq:RCU_OMA_SFP}, $P_{K_\s}(k_\s) \!=\! {\rm Bino}(k_\s; \sf{K},\rho_\s)$ and $\nu_1 \!\triangleq\! 2 - \sum_{k = k_{\s,\ell}}^{k_{\s,u}} \!P_{K_\s}(k) - \E[{K}_\s]{\frac{\sf{M}_\s!}{\sf{M}_\s^{{K}_\s}(\sf{M}_\s-{K}_\s)!} } + \sf{K} \rho_\s  \frac{\Gamma(\sf{n}_\s/2,\sf{n}_\s \sf{P}_\s/(2\sf{P}'_\s))}{\Gamma(\sf{n}_\s/2)}$. In~\eqref{eq:bar_eps_SMD} and~\eqref{eq:bar_eps_SFP}, we define $p_t \triangleq \sum_{t'\in \overline{\Tc}_t} p_{t,t'}$ and $p_{t,t'} \triangleq e^{-\sf{n}_\s E(t,t')/2}$ where
		\begin{align}
			E(t,t') &\triangleq \max_{\rho_1,\rho_2 \in [0,1]} -\rho_1\rho_2 t' R_1 - \rho_2 R_2 + E_0(\rho_1,\rho_2), \label{eq:Ett} \\
			E_0(\rho_1,\rho_2) &\triangleq \max_{\lambda} \rho_2 a + \ln\big[1-\rho_2 (1  + \big((k_\s - \overline{k_\s'})^+ \notag \\
            &\qquad + (\underline{k_\s'} - k_\s)^+)\sf{P}'_\s\big) b\big], \label{eq:E0}
        \end{align}
        with $a \triangleq \rho_1 \ln(1+ \sf{P}_\s' t' \lambda) + \ln(1+ \sf{P}_\s't \mu)$, 
		$b \triangleq \rho_1\lambda -\frac{\mu}{1+ \sf{P}'_\s t\mu}$, 
		$\mu \triangleq \frac{\rho _1\lambda}{1+\sf{P}_\s't'\lambda}$, $R_1 \triangleq  \frac{2}{\sf{n}_\s t'} \ln\binom{\sf{M}_\s - \max\{k_\s,\underline{k_\s'}\}}{t'}$, and $R_2 \triangleq  \frac{2}{\sf{n}_\s} \ln \binom{\min\{k_\s, \overline{k_\s'}\}}{t}$.
        We also define
        \begin{align}
			q_t &\triangleq \inf_{\gamma} \big(\P{{I}_{t} \le \gamma} + \textstyle\sum_{t'\in \overline{\Tc}_t}
			\exp(\sf{n}_\s(t'R_1 + R_2)/2 - \gamma)\big), \label{eq:qt}\\
			q_{t,t'} &\triangleq \inf_{\gamma} \big(\P{{I}_{t} \le \gamma} + \exp(\sf{n}_\s(t'R_1 + R_2)/2 - \gamma)\big), \label{eq:qtt}
        \end{align}
        with 
					\begin{equation} \label{eq:def_It}
			{I}_t \triangleq \min_{\Wc'_{2} \subset [(k_\s - \overline{k_\s'})^+ + 1:k_\s], \atop |\Wc_{2}'| = t} \imath_t(c(\Wc_{1}') + c(\Wc'_{2});{\vec y}_\s  \big\vert c([k_\s] \setminus \Wc')),
							\end{equation} 
		where $\Wc_{1}' = [k_\s + 1: \underline{k_\s'}]$, $\Wc' = [(k_\s - \overline{k_\s'})^+] \cup \Wc'_{2}$, 
		and 
			$\imath_t(c(\Wc');{\vec y}_\s \vert c(\Wc \setminus \Wc')) 
			\triangleq \frac{\sf{n}_\s}{2}\ln(1+(t+(k_\s-\overline{k_\s'})^+)\sf{P}_\s')  + \frac{1}{2}\big(\frac{\|{\vec y}_\s - c(\Wc \setminus \Wc')\|^2}{1+(t+(k_\s-\overline{k_\s'})^+)\sf{P}'_\s} 
			\!-\! \|{\vec y}_\s \!-\! c(\Wc') \!-\! c(\Wc \!\setminus\! \Wc')\|^2\big)$.  
        Furthermore, we define the sets $\Tc$, $\Tc_t$, and $\overline{\Tc}_t$ as
        \begin{align}
			\Tc &\triangleq [0:\min\{\overline{k_\s'},k_\s,\sf{M}_\s-\underline{k_\s'} - (k_\s - \overline{k_\s'})^+\}], \label{eq:T} \\
			\Tc_t &\triangleq \big[\big({(k_\s - \overline{k_\s'})}^+ - {(\underline{k_\s'} - k_\s)}^+ + \max\{\underline{k_\s'},1\} \notag \\ 		& \qquad \big. \big. 
			- k_\s + t\big)^+ : u_t\big],	\label{eq:Tt}	\\
			\overline{\Tc}_t &\triangleq \big[\big({(k_\s - \overline{k_\s'})}^+ - {(k_\s-\underline{k_\s'})}^+ + t\big)^+ : u_t \big], \label{eq:Tbart}
        \end{align}
        with $u_t \triangleq \min\big\{{(\overline{k_\s'} - k_\s)}^+ - {(\underline{k_\s'}-k_\s)}^+ + t,  
			\overline{k_\s'} - {(\underline{k_\s'}-k_\s)}^+, \sf{M}_\s-\max\{\underline{k_\s'},k_\s\}\big\}.$
        Finally, 
        \begin{equation}
            \xi(k_\s,k_\s') \triangleq 
			\min_{k \in [0:\sf{K}]\setminus \{k_\s'\}} \P{m_\s\left({\vec Y}_\s',k_\s' \right) > m_\s\left({\vec Y}_\s',k\right)} \label{eq:xi}
		\end{equation}}
        with ${\vec Y}_\s' \sim \Nc(\mathbf{0},(1+k_\s P')\vec I _{\sf{n}_\s})$.
	\end{theo}
	\begin{IEEEproof}
		We obtain $\epsilon_{\smd}$ and $\epsilon_{\sfp}$ directly by adapting 
		\cite[Th.~1]{Ngo2022} to the real-valued case. 
        \revisee{Here, $\nu_1$ is obtained from a change of measure under which  $k_{\s,\ell} \le K_\s \le k_{\s,u}$,  $\vec X_{\s,k} = \vec C_{\s,W_{\s,k}}$ (instead of $\vec X_{\s,k} = \vec C_{\s,W_{\s,k}}\ind{\|\vec C_{\s,W_{\s,k}}\|^2 \le \sf{n}_\s\sf{P}_\s}$), and the standard users transmit distinct codewords. The quantities $p_{t,t'}$ and $q_{t,t'}$ are upper bounds on the probability of having $t + (k_\s - \overline{k_\s'})^+$ SMDs and $t' + (\underline{k_\s'} - k_\s)^+$ SFPs;  $p_{t}$ and $q_t$ are upper bounds on the probability of having $t + (k_\s - \overline{k_\s'})^+$ SMDs. We obtain $p_t$ and $p_{t,t'}$ based on an error-exponent analysis, while $q_t$ and $q_{t,t'}$ follow from a variation of the dependence-testing bound~\cite[Th.~17]{Polyanskiy2010}. The sets $\mathcal{T}$, $\mathcal{T}_t$, $\overline{\mathcal{T}}_t$ contain possible values of $t$ and $t'$. Finally, $\xi(k_\s,k_\s')$ is an upper bound on the probability that, given $K_\s = k_\s$ standard users, the estimation step~\eqref{eq:Ks_estimation} returns $k'_\s$.}
	\end{IEEEproof}
 
		The limits $k_{\s,\ell}$ and $k_{\s,u}$ are introduced to facilitate the numerical evaluation of the bounds, since they allow \revise{us} to avoid large sums (from $0$ to $\sf{K}$) over $k_\s$ and $k'_\s$. We set $k_{\s,\ell}$ to be the largest value and $k_{\s,u}$ the smallest value for which $\sum_{k = k_{\s,\ell}}^{k_{\s,u}} P_{K_\s}(k)$ exceeds a given threshold.
 
	\begin{remark} \label{rem:std_dec_rad}
		As explained in~\cite{Ngo2021ISITmassive,Ngo2022}, when $\sf{P}_\s$ is small,  one should use a small $r_\s$ to avoid noise overfitting. Specifically, when the noise dominates, a large $r_\s$ seems to increase the chance that the decoding step~\eqref{eq:std_decoding} returns a list containing codewords whose sum is closer in Euclidean distance to the noise than to the sum of the transmitted codewords. To satisfy mild targets on $\epsilon_{\smd}$ and $\epsilon_{\sfp}$, setting $r_\s = 0$ 
        results in higher energy efficiency than $r_\s > 0$, see~\cite[Fig.~2]{Ngo2022}.
	\end{remark}
	
	\subsubsection{Overall Random-Coding Bound}
	By combining 
	Theorem~\ref{th:alarm_block} and 
	Theorem~\ref{th:std_block}, we obtain the following random-coding bound for \gls{HOMA}.
	
	\begin{theo}[Random-coding bound for \gls{HOMA}] \label{th:RCU_HOMA}
		Fix $r_\a$, $r_\s$, $\sf{n}_{\a} \in [0:\sf{n}]$, $k_{\a, \ell} \in [1:\sf{K}]$, $k_{\a, u} \in [k_{\a, \ell} + 1:\sf{K}]$, $k_{\s, \ell} \in [0:\sf{K}]$, $k_{\s, u} \in [k_{\s, \ell} + 1:\sf{K}]$, $(\sf{P}_\a,\sf{P}_\s)$ such that $\sf{n}_\a \sf{P}_\a + \sf{n}_\s \sf{P}_\s \le \sf{n} \sf{P}$, $\sf{P}_\a' < \sf{P}_\a$, and $\sf{P}_\s' < \sf{P}_\s$. For the considered Gaussian \gls{MAC} with both standard and alarm traffic, there exists an $(\sf{M}_\a,\sf{M}_\s,\sf{n},\epsilon_\amd,\epsilon_\afp,\epsilon_\smd,\epsilon_\sfp)$ random-access code with $\epsilon_\amd$ and $\epsilon_\afp$ given in Theorem~\ref{th:alarm_block}, and $\epsilon_\smd$ and $\epsilon_\sfp$ given in Theorem~\ref{th:std_block}.
	\end{theo}
	
	\section{Heterogeneous Nonorthogonal Multiple Access}\label{sec:HNOMA}
	We consider now a nonorthogonal network slicing strategy, which we refer to as \gls{HNOMA}, where both standard and alarm codewords are transmitted \revise{over} the whole frame. 
    \revise{We illustrate a frame of \gls{HNOMA} in Fig.~\ref{fig:HNOMA}.}

            

	\subsection{Signal Model}
    Each alarm user maps the alarm message $W_{\a}$ to an alarm codeword $\vec X_\a \in \RR^{\sf{n}}$. Furthermore, the $k$th standard user maps its standard message $W_{\s,k}$ to a standard codeword $\vec X_{\s,k} \in \RR^{\sf{n}}$. Over a frame, user~$k$ transmits $\vec S_k = \vec X_\a + \vec X_{\s,k}$ if it has both a standard message and an alarm message, $\vec S_k = \vec X_\a$ if it has only an alarm message, $\vec S_k = \vec X_{\s,k}$ if it has only a standard message, and $\vec S_k = \mathbf{0}$ if it is inactive. The received signal~\eqref{eq:model} can be written as 
    \begin{equation}
    \vec Y = K_\a \vec X_\a + \sum_{k = 1}^{K_\s} \vec X_{\s,k} + \vec Z,
    \end{equation}
    where $K_\a$ is the number of alarm users, $K_\s$ is the number of standard users, and we assume that the first $K_\s$ users are the standard users. As for~\gls{HOMA}, $K_\a = 0$ if no alarm event occurs and $K_\a \sim {\rm Bino}(\sf{K},\rho_\d)$ otherwise; furthermore, $K_\s \sim {\rm Bino}(\sf{K},\rho_\s)$. We impose the power constraints $\|\vec X_\a\|^2/\sf{n} \le \sf{P}_\a$ and $\|\vec X_{\s,k}\|^2/\sf{n} \le \sf{P}_\s$, $k \in [K_\s]$, and set $\sf{P}_\a + \sf{P}_\s \le \sf{P}$ to satisfy the overall power constraint. The average energy per bit of alarm and standard traffic are upper-bounded by $(\EbNo)_{\a} =
	\frac{\sf{n} \sf{P}_\a \rho_\d \sf{K}}{2\log_{2}\sf{M}_\a}$ and $(\EbNo)_{\s} =
	\frac{\sf{n} \sf{P}_\s}{2\log_{2}\sf{M}_\s}$, respectively.

	\subsection{Random-Coding Bound}
	We fix \revise{the transmit power} $\sf{P}_\a' < \sf{P}_\a$ and draw the alarm codewords $\Cc_\a = \{\vec C_{\a,1},\dots,\vec C_{\a, \sf{M}_\a}\}$ independently from $\Nc(\mathbf{0}, \sf{P}_\a' \vec I_{\sf{n}})$. Similarly, we fix \revise{the transmit power} $\sf{P}_\s' < \sf{P}_\s$ and draw the standard codewords $\Cc_\s = \{\vec C_{\s, 1},\dots,\vec C_{\s, \sf{M}_\s}\}$ independently from $\Nc(\mathbf{0}, \sf{P}_\s' \vec I_{\sf{n}})$. The alarm message $W_\a$ is mapped to $\vec X_\a = \vec C_{\a, W_\a} \mathbbm{1}\big\{\|\vec C_{\a, W_\a}\|^2 \le \sf{n} \sf{P}_\a\big\}$, while the standard message $W_{\s,k}$ is mapped to $\vec X_{\s,k} = \vec C_{\s, W_{\s,k}} \ind{\|\vec C_{\s, W_{\s,k}}\|^2 \le \sf{n} \sf{P}_\s}$. To convey $(W_\a,W_{\s,k})$, $W_\a$, or $W_{\s,k}$, user~$k$ transmits $\vec X_\a + \vec X_{\s,k}$, $\vec X_\a$, or $\vec X_{\s,k}$, respectively. 
	
	Given a realization $\vec y$ of the received signal, the receiver first decodes the alarm message and estimates the number of alarm users similarly to the alarm block in \gls{HOMA}. The sum of the standard codewords, $\sum_{k = 1}^{K_\s} \vec X_{\s,k}$, is treated as noise. Specifically, the receiver performs steps~\eqref{eq:Ka_estimation} and~\eqref{eq:alarm_decoding} with $\vec y_\a$ replaced by $\vec y$, and obtains an estimate $(\widehat{K}_\a, \widehat{W}_{\a})$ of $(K_\a,W_\a)$. Next, exploiting reliability diversity~\cite{Popovski2018}, the receiver performs interference cancellation and decodes the list of standard messages in a similar manner to the standard block in~\gls{HOMA}. Specifically, the receiver removes the decoded alarm codeword from the received signal to obtain $\vec y_{\ic} = \vec y-\widehat K_{\a}\vec C_{\a, \widehat W_\a}$. It then performs steps~\eqref{eq:Ks_estimation} and~\eqref{eq:std_decoding} with $\vec y_\s$ replaced by $\vec y_{\ic}$, and obtains an estimate $\widehat{\Wc}_\s$ of the list of transmitted standard messages. 
	
	An error analysis of the proposed scheme 
	leads to the following random-coding bound.
	\begin{theo}[Random-coding bound for \gls{HNOMA}] \label{th:HNOMA}
		Fix $r_\a$, $r_\s$, $k_{\a, \ell} \in [1:\sf{K}]$, $k_{\a, u} \in [k_{\a, \ell} + 1:\sf{K}]$, $k_{\s, \ell} \in [0:\sf{K}]$, $k_{\s, u} \in [k_{\s, \ell} + 1:\sf{K}]$, $(\sf{P}_\a,\sf{P}_\s)$ such that $\sf{P}_\a + \sf{P}_\s \le \sf{P}$, $\sf{P}_\a' < \sf{P}_\a$, and $\sf{P}_\s' < \sf{P}_\s$. For the considered Gaussian \gls{MAC} with both standard and alarm traffic, there exists an $(\sf{M}_\a,\sf{M}_\s,\sf{n},\epsilon_\amd,\epsilon_\afp,\epsilon_\smd,\epsilon_\sfp)$ random-access code satisfying the power constraint $\sf{P}$ for which 
		$\epsilon_\amd = \sum_{{k}_\s = {k}_{\s,\ell}}^{{k}_{\s,u}} P_{{K}_\s}( k_\s) \bar{\epsilon}_\amd(k_\s) + \nu_2$, $\epsilon_\afp =  \sum_{k_\s = k_{\s,\ell}}^{k_{\s,u}} P_{{K}_\s}(k_\s) \bar{\epsilon}_\afp(k_\s) + \nu_3$, $\epsilon_\smd = \max\{\epsilon_{\smd \vert \Ac}, \epsilon_{\smd | \bar{\Ac}}\}$, and
		$\epsilon_\sfp = \max\{\epsilon_{\sfp \vert \Ac}, \epsilon_{\sfp | \bar{\Ac}}\}$. Here, 
		\begin{align}
			\nu_2 &\triangleq \frac{\Gamma\big(\frac{\sf{n}}{2},\frac{\sf{n} \sf{P}_\a}{2\sf{P}'_\a}\big)}{\Gamma(\sf{n}/2)} + 1 - \sum_{k = k_{\a,\ell}}^{k_{\a,u}} P_{K_\a}(k) + \nu_3,  \label{eq:p2}\\ 
			\nu_3 &\triangleq \sf{K}\rho_\s \frac{\Gamma\big(\frac{\sf{n}}{2},\frac{\sf{n} \sf{P}_\s}{2\sf{P}'_\s}\big)}{\Gamma(\sf{n}/2)} + 2 - \sum_{k = {k}_{\s,\ell}}^{{k}_{\s,u}} P_{{K}_\s}(k) \notag \\
            &\quad - \E[{K}_\s]{\frac{\sf{M}_\s}{\sf{M}_\s^{{K}_\s} (\sf{M}_\s-{K}_\s)!}}, \label{eq:p3} \\ 
			\bar{\epsilon}_\amd(k_\s) &\triangleq \sum_{k_\a = k_{\a,\ell}}^{k_{\a,u}} P_{K_\a}(k_\a) \Bigg(\eta(k_\a,0,k_\s) \notag \\
			&\qquad + \min\Bigg\{1,\sum_{k'_\a=k_{\a,\ell}}^{k_{\a,u}} \eta(k_{\a},{k}'_{\a},  k_\s)\Bigg\} \notag \\
            &\qquad\quad \cdot \sum_{\hat{k}_\a \in \{0\} \cup [k_{\a,\ell} :k_{\a,u}]} \theta_\amd(k_\a,\hat k_\a, k_\s) \Bigg), \label{eq:bar_eps_AMD}\\ 
			\bar{\epsilon}_\afp(k_\s) &\triangleq \sum_{\hat{k}_{\a} = k_{\a,\ell}}^{k_{\a,u}} \min\Bigg\{1,\sum_{k_\a' = k_{\a,\ell}}^{k_{\a,u}} \eta(0, {k}_\a',k_\s) \Bigg\}  \theta_\afp(\hat{k}_\a,k_\s), \label{eq:bar_eps_AFP} 
		\end{align}
		where $P_{{K}_\s}( k_\s) 
		= {\rm Bino}( k_\s;\sf{K},\rho_\s)$ and $P_{K_\a}(k_\a) = {\rm Bino}(k_\a;\sf{K},\rho_\d)$. We define $\theta_\amd(k_\a,\hat k_\a, k_\s)$ similarly to $\gamma_\amd(k_\a,\hat k_\a)$ in~\eqref{eq:gamma_md} except that $\sf{n}_\a$ is replaced by $\sf{n}$, and given $X'_i = x'_i$, we have that $Y'_i \sim \mathcal{N}({k}_{\a}x'_i,1+{k}_\s \sf{P}'_\s)$ instead of $Y'_i \sim \mathcal{N}({k}_{\a}x'_i,1)$. We define $\theta_\afp(\hat{k}_\a,k_\s)$ similarly to $\gamma_\afp(\hat{k}_\a)$ in~\eqref{eq:gamma_fp} except that $\sf{n}_\a$ is replaced by $\sf{n}$ and $\beta$ is given by $s(1 - (1+2\hat{k}_{\a}^2\sf{P}'_\a s)^{-1} ) (1+k_\s \sf{P}'_\s)$. Furthermore, we define $\eta(k_{\a},k'_{\a}, k_\s)$ similarly to $\zeta(k_\a,k_\a')$ in~\eqref{eq:zeta}, except that $\vec Y' \sim \mathcal{N}(\mathbf{0}, (1+k_{\a}^2 \sf{P}'_\a + {k}_\s \sf{P}'_\s)\vec I_{\sf{n}})$. 
		We also have that
		\begin{align}
			\epsilon_{\smd | \bar{\Ac}} &\triangleq 1 - \sum_{{k}_\s = {k}_{\s,\ell}}^{{k}_{\s,u}} P_{{K}_\s}( k_\s) (1-\bar{\epsilon}_\afp(k_\s)) \notag \\
            &\qquad \quad \cdot (1-\bar{\epsilon}_\smd(\sf{M}_\s,\sf{n},k_\s,\sf{P}_\s,\sf{P}'_\s,r_\s,k_{\s,\ell},k_{\s,u})) \notag \\
            &\quad + \nu_2, \label{eq:RCU_NOMA_SMD_noAlarm}\\
			\epsilon_{\sfp | \bar{\Ac}} &\triangleq 1 - \sum_{{k}_\s = {k}_{\s,\ell}}^{{k}_{\s,u}} P_{{K}_\s}( k_\s) (1-\bar{\epsilon}_\afp(k_\s)) \notag \\
            &\quad \qquad \cdot (1-\bar{\epsilon}_\sfp(\sf{M}_\s,\sf{n},k_\s,\sf{P}_\s,\sf{P}'_\s,r_\s,k_{\s,\ell},k_{\s,u})) \notag \\
            &\quad + \nu_2, \label{eq:RCU_NOMA_SFP_noAlarm}\\ 
			\epsilon_{\smd \vert \Ac} &\triangleq 1 - \sum_{{k}_\s = {k}_{\s,\ell}}^{{k}_{\s,u}} P_{{K}_\s}( k_\s) (1-\bar{\epsilon}_\amd(k_\s)) \notag \\
            &\quad \cdot  \Bigg( 1- \sum_{k_\a = k_{\a,\ell} }^{k_{\a,u}} P_{K_\a}(k_\a) \sum_{\hat{k}_\a \in \{0\} \cup [k_{\a,\ell}:k_{\a,u}]} \!\!\alpha(k_\a,\hat{k}_\a,k_\s)
			\notag \\
			&\qquad \quad \cdot \bar{\epsilon}_\smd\big(\sf{M}_\s,\sf{n},k_\s,\tfrac{\sf{P}_\s}{\revise{1 + (k_\a - \hat{k}_\a)^2 \sf{P}_\a'}},\tfrac{\sf{P}'_\s}{\revise{1 + (k_\a - \hat{k}_\a)^2 \sf{P}_\a'}}, \notag \\
            &\qquad\qquad\qquad  r_\s,k_{\s,\ell},k_{\s,u}\big)  \Bigg) + \nu_2, \label{eq:RCU_NOMA_SMD_alarm}\\ 
			\epsilon_{\sfp \vert \Ac} &\triangleq 1 - \sum_{{k}_\s = {k}_{\s,\ell}}^{{k}_{\s,u}} P_{{K}_\s}( k_\s) (1-\bar{\epsilon}_\amd(k_\s))  \notag \\
            &\quad\cdot \Bigg( 1- \sum_{k_\a = k_{\a,\ell} }^{k_{\a,u}} P_{K_\a}(k_\a) \sum_{\hat{k}_\a \in \{0\} \cup [k_{\a,\ell}:k_{\a,u}]} \!\!\alpha(k_\a,\hat{k}_\a,k_\s)
			\notag \\
			&\qquad \quad \cdot  \bar{\epsilon}_\sfp\big(\sf{M}_\s,\sf{n},k_\s,\tfrac{\sf{P}_\s}{\revise{1 + (k_\a - \hat{k}_\a)^2 \sf{P}_\a'}},\tfrac{\sf{P}'_\s}{\revise{1 + (k_\a - \hat{k}_\a)^2 \sf{P}_\a'}}, \notag \\
            &\qquad \qquad \qquad r_\s,k_{\s,\ell},k_{\s,u}\big)  \Bigg) + \nu_2, \label{eq:RCU_NOMA_SFP_alarm}
		\end{align}
		where $\bar{\epsilon}_\smd$ and $\bar{\epsilon}_\sfp$ are given in~\eqref{eq:bar_eps_SMD} and~\eqref{eq:bar_eps_SFP}, respectively, 
		and
			\begin{align}
				&\alpha(k_\a,\hat{k}_\a,k_\s) \triangleq \ind{\hat{k}_\a = 0}\eta(k_\a,0,k_\s) \notag \\ \quad &+  \min\bigg\{1,\sum_{k_\a' = k_{\a,\ell}}^{k_{\a,u}} \eta(k_\a, k'_\a,k_\s)\bigg\} \bigg(1\!+\!\frac{(k_\a - \hat{k}_\a)^2 \sf{P}'_\a}{4(1+\bar{k}_\s \sf{P}'_\s)}\bigg)^{-\sf{n}/2}.
			\end{align}
		\end{theo}
		\begin{IEEEproof}
			The proof follows by adapting the bounding techniques used for \gls{HOMA} to account for the interference from the standard codewords when decoding the alarm message, and the residual interference from the alarm codeword when decoding the standard messages, induced by the interference cancellation process in \gls{HNOMA}. 
            See Appendix~\ref{proof:HNOMA} for details.
		\end{IEEEproof}
		
					The term $\alpha(k_\a,\hat{k}_\a,k_\s)$ is an upper bound on the probability that the receiver estimates $K_\a$ by $\hat{k}_\a$, given that there are $k_\a$ alarm users, $k_\s$ standard users, and 
     $W_\a$ is correctly decoded. 
		\section{Numerical Experiments} \label{sec:numerical}
		We set the framelength $\sf{n}$ to $30000$. Motivated by Remark~\ref{rem:std_vs_alarm}, we consider $(\sf{M}_\s, \sf{M}_\a) = (2^{100},2^3)$, the mild target reliability $\max\{P_\smd, P_\sfp\} \le 10^{-1}$ for the standard traffic, and the stringent target reliability $\max\{P_\amd, P_\afp\} \le 10^{-5}$ for the alarm traffic. We set $\sf{K} \in [1000:30000]$ and $\rho_\s = 0.01$, so that $\E{K_\s} \in [10 : 300]$, similar to the setting in~\cite{Polyanskiy2017, Ngo2022}. 
        Let $(\EbNo)_\s^*$ be the minimum required energy per bit for the standard traffic \revise{to satisfy $\max\{P_\smd, P_\sfp\} \le 10^{-1}$} if the alarm traffic is not present. We evaluate $(\EbNo)_\s^*$ in a similar manner as in~\cite{Ngo2022}, considering maximum-likelihood estimation of $K_\s$ and zero standard-message decoding radius, i.e., $r_\s = 0$. Note that for the mild requirement $\max\{P_\smd, P_\sfp\} \le 10^{-1}$, setting $r_\s = 0$ leads to high energy efficiency since it helps to avoid noise overfitting; see Remark~\ref{rem:std_dec_rad}. In the remainder of this section, we address the following question: {\em Let the standard traffic operate at $(\EbNo)_\s^* + \delta$~(dB) for a fixed backoff $\delta > 0$. What is the minimum required\footnote{
        \revise{Report $(\EbNo)_\a$ for a fixed $(\EbNo)_\s$ is meaningful because it reflects the energy efficiency of the alarm traffic even if the alarm event is rare.}} $(\EbNo)_\a$?
        }
		
		\subsection{\gls{HOMA}}
\begin{figure}[t]
			\centering
			\begin{tikzpicture}
				\tikzstyle{every node}=[font=\footnotesize]
				\begin{axis}[%
					width=3in,
					height=1.7in,
					at={(0.759in,0.481in)},
					scale only axis,
					xmin=0,
					xmax=30000,
					xlabel style={font=\color{black},yshift=1ex},
					xlabel={\footnotesize Total number of users $\sf{K}$},
					ymin=0,
					ymax=20000,
					ylabel style={font=\color{black}, yshift=-2ex},
					ylabel={\footnotesize $n_{\a,\max}$},
					axis background/.style={fill=white},
					xmajorgrids,
					ymajorgrids,
					legend style={at={(0.99,0.99)}, anchor=north east, row sep=-2.5pt, legend cell align=left, align=left, draw=white!15!black, nodes={scale=0.95},fill=white, fill opacity=0.8, text opacity=1, draw=none}
					]
					
					\addplot [color=red,line width=1pt]
					table[row sep=crcr]{%
						1000       22642 \\
						2500       17484 \\
						5000        5152 \\
						10000         468 \\
						15000          28 \\
						20000           2 \\
						25000           2 \\
						30000           2 \\
					};
					\addlegendentry{$\delta = 0$~dB};
					
					\addplot [color=red,dashed,line width=1pt]
					table[row sep=crcr]{%
						1000       22760 \\
						2500       19098 \\
						5000       14762 \\
						10000        7584 \\
						15000        1932 \\
						20000         964 \\
						25000         694 \\
						30000         526 \\
					};
					\addlegendentry{$\delta = 0.1$~dB};
					
					\addplot [color=red,dashdotted,line width=1pt]
					table[row sep=crcr]{%
						1000       22840 \\
						2500       19842 \\
						5000       16538 \\
						10000        9296 \\
						15000        3306 \\
						20000        1850 \\
						25000        1344 \\
						30000        1022 \\
					};
					\addlegendentry{$\delta = 0.2$~dB};
				\end{axis}
			\end{tikzpicture}%
			\caption{The number of available channel uses $n_{\a,\max}$ for the alarm traffic in~\gls{HOMA} when the standard traffic operates at $(\EbNo)_\s^* + \delta$~dB and satisfies  $\max\{\epsilon_\smd, \epsilon_\sfp\} \le 10^{-1}$. Here, $\sf{n} = 30000$, $\sf{M}_\s = 2^{100}$ and $\rho_\s = 0.01$. 
			}
			\label{fig:ns_backoff}
		\end{figure}
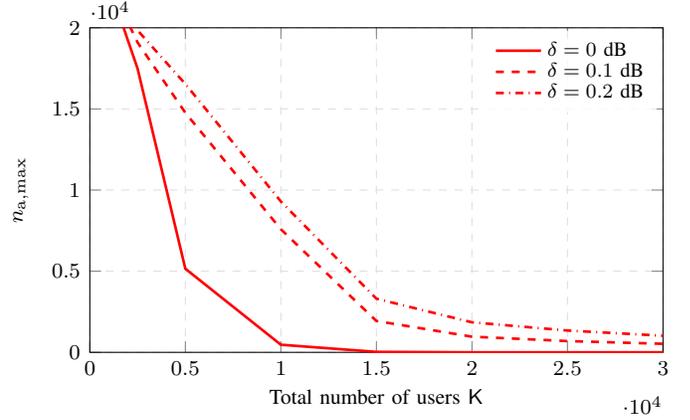

		To address this question for the case of \gls{HOMA}, we find the minimum blocklength $\sf{n}_{\s,\min}$ required to satisfy
		$\max\{\epsilon_\smd, \epsilon_\sfp\} \le 10^{-1}$ at $(\EbNo)_\s^* + \delta$~dB. The number of available channel uses for the alarm traffic is thus $\sf{n}_{\a,\max} = \sf{n} - \sf{n}_{\s,\min}$. 
		In Fig.~\ref{fig:ns_backoff}, we plot $\sf{n}_{\a,\max}$ as a function of $\sf{K}$ for three backoff values: $0$~dB (no backoff), $0.1$~dB, and $0.2$~dB. We see that $\sf{n}_{\a,\max}$ is large for small $\sf{K}$, and then decreases as $\sf{K}$ becomes large. For $K \le 15000$, even with zero backoff, $\sf{n}_{\s,\min}$ is less than $\sf{n}$. This shows that for a fixed $(\EbNo)_\s$, i.e., a fixed energy $\sf{n}_\s \sf{P}_\s$, the higher transmit power $\sf{P}_\s$ induced by reducing $\sf{n}_\s$ is sufficient to keep $\max\{\epsilon_\smd, \epsilon_\sfp\}$  below $10^{-1}$. For $K > 15000$, the standard blocklength cannot be shortened without a positive backoff, as a higher power is needed to counteract multi-user interference. As expected, a larger backoff leads to more available channel uses for the alarm traffic.
		
		Next, for a given $n_{\a,\max}$, \revise{we find the minimum required $(\EbNo)_\a$ by solving 
        \begin{align}
            \minimize_{\sf{n}_\a \in [\sf{n}_{\a,\max}], \atop \rho_{\d} \in (1 - (10^{-5})^{1/K},\rho_{\d,\max}), \; \sf{P}_\a > 0, \sf{P}_\a' \le \sf{P}_\a} &(\EbNo)_\a \label{eq:minEbNoa}\\
            \text{subject to} \qquad \qquad \qquad&\max\{\epsilon_\amd, \epsilon_\afp\} \le 10^{-5} \notag
        \end{align}
        with $\epsilon_\amd$ and $\epsilon_\afp$ given in Theorem~\ref{th:alarm_block}.
        To solve~\eqref{eq:minEbNoa}, we apply a golden-section search sequentially over $\sf{n}_\a$ and $\rho_{\d}$, where for each value of $\sf{n}_\a$ and $\rho_\d$, the required $(\EbNo)_\a$ is obtained via a binary search for the smallest power $\sf{P}_\a$ such that $\max\{\epsilon_\amd, \epsilon_\afp\} \le 10^{-5}$. For each $\sf{n}_\a$ and $\sf{P}_\a$, we choose the transmit power $\sf{P}'_\a$ such that $\frac{\Gamma(\sf{n}_\a/2,\sf{n}_\a \sf{P}_\a/(2\sf{P}'_\a))}{\Gamma(\sf{n}_\a/2)} < 10^{-8}$ to limit $\nu_0$.} We choose $k_{\a,\ell}$ and $k_{\a,u}$ such that $k_{\a,\ell} \ge 2$ and  $\P{K_\a \notin [k_{\a,\ell}: k_{\a,u} ]\vert \Ac} = 1 - \sum_{k = k_{\a,\ell}}^{k_{\a,u}}{\rm Bino}(k;\sf{K},\rho_\d) < 10^{-10}$.  
		In the following, we set $\delta = 0.1$~dB and study the minimum required $(\EbNo)_\a$.
		\subsubsection{Impact of the Device Sensitivity}
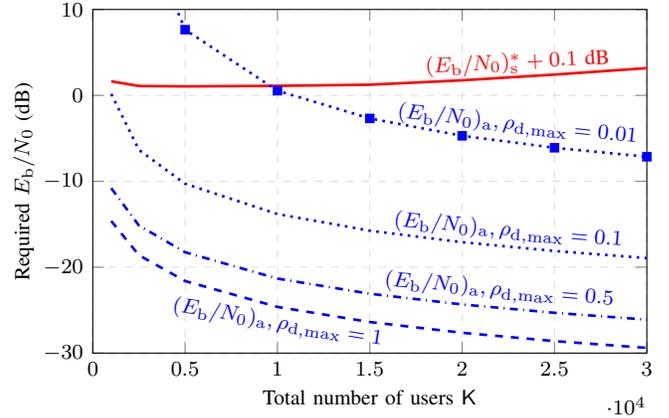
\begin{figure}[t]
			\centering
			\begin{tikzpicture}
				\tikzstyle{every node}=[font=\footnotesize]
				\begin{axis}[%
					width=2.9in,
					height=1.8in,
					at={(0.759in,0.481in)},
					scale only axis,
					xmin=0,
					xmax=30000,
					xlabel style={font=\color{black},yshift=1ex},
					xlabel={\footnotesize Total number of users $\sf{K}$},
					ymin=-30,
					ymax=10,
					ylabel style={font=\color{black}, yshift=-2ex},
					ylabel={\footnotesize Required $E_{\rm b}/N_0$ (dB)},
					axis background/.style={fill=white},
					xmajorgrids,
					ymajorgrids,
					legend style={at={(0.99,0.32)}, anchor=south east, row sep=-2.5pt, legend cell align=left, align=left, draw=white!15!black, nodes={scale=0.92},fill=white, fill opacity=0.2, text opacity=1, draw=none}
					]
					\addplot [color=red,line width=1pt]
					table[row sep=crcr]{
						1000 1.638818835868710 \\ 
						2500 1.093003773659495 \\ 
						5000 1.046901113925378 \\ 
						10000 1.108300355924746 \\ 
						15000 1.241000190960597 \\ 
						20000 1.750954301244553 \\ 
						25000 2.419138257681345 \\ 
						30000 3.174586008254130 \\ 
					};
					\node[rotate=3,color=red] at (axis cs:2.3e4,3.8) () {$(\EbNo)_{\s}^* + 0.1$~dB};
					
					\addplot [color=blue,dashed,line width=1pt]
					table[row sep=crcr]{%
						1000 -14.622982195687996 \\
						2500  -18.602333375364935 \\
						5000 -21.612451758486770 \\
						10000 -24.622988972511735 \\
						15000 -26.383838627586272 \\
						20000 -27.632562709228864 \\
						25000 -28.601399885797377 \\
						30000  -29.392620668723794 \\
					};
					\node[rotate=-9,color=blue] at (axis cs:1e4,-26.5) () {$(\EbNo)_{\a}, \rho_{\d,\max} = 1$};
					
					\addplot [color=blue,dashdotted,line width=1pt]
					table[row sep=crcr]{%
						1000 -10.802828710828976 \\
						2500  -15.150679238684461 \\
						5000 -18.261656113701918  \\
						10000 -21.325379702428961 \\
						15000 -23.089473708032642 \\
						20000 -24.336421034569639 \\
						25000 -25.3071 \\
						30000 -26.1024  \\
					};
					\node[rotate=-5,color=blue] at (axis cs:2.2e4,-22.7) () {$(\EbNo)_{\a}, \rho_{\d,\max} = 0.5$};
					
					\addplot [color=blue,dotted,line width=1pt]
					table[row sep=crcr]{%
						1000 0.171689382995282 \\
						2500  -6.346910588341586 \\
						5000 -10.288916440596552 \\
						10000 -13.815686257570860 \\
						15000 -15.751375171298264 \\
						20000 -17.092353439225462 \\
						25000 -18.111636955772724 \\
						30000  -18.945547890963383 \\
					};
					\node[rotate=-5,color=blue] at (axis cs:2.25e4,-15.8) () {$(\EbNo)_{\a}, \rho_{\d,\max} = 0.1$};
					
					\addplot [color=blue,dotted,mark=square*,mark size=1.3,line width=1pt,mark options={solid}]
					table[row sep=crcr]{%
						2500 20.675268211488401 \\
						5000 7.6515 \\
						10000 0.5438 \\
						15000 -2.6822 \\
						20000 -4.7056 \\
						25000 -6.0961 \\
						30000  -7.1446 \\
					};
					\node[rotate=-6,color=blue] at (axis cs:2.3e4,-3.5) () {$(\EbNo)_{\a}, \rho_{\d,\max} = 0.01$};
				\end{axis}
			\end{tikzpicture}%
			\caption{The minimum  $(\EbNo)_\a$ required for \gls{HOMA} to satisfy $\max\{\epsilon_\amd, \epsilon_\afp\} \le 10^{-5}$ when $(\EbNo)_{\s}= (\EbNo)_{\s}^* + 0.1$~dB for different values of the device sensitivity $\rho_{\d,\max}$. Here, $\sf{n} = 30000$, $\sf{M}_\a = 2^3$, $\sf{M}_\s = 2^{100}$, and $\rho_\s = 0.01$.} 
			\label{fig:EbN0_sensitivity}
		\end{figure}

		As noted in Remark~\ref{rem:rho_d_max}, $\rho_\d$ 
		is upper-bounded by the probability that a user detects the alarm event, $\rho_{\d,\max}$.  
		To study the impact of $\rho_{\d,\max}$ on the alarm-traffic energy efficiency,
		we vary its value and show in Fig.~\ref{fig:EbN0_sensitivity} the corresponding minimum required $(\EbNo)_\a$ as a function of $\sf{K}$. We also show the value of  $(\EbNo)_{\s}^* + 0.1$~dB for reference. We see that the required $(\EbNo)_\a$ can be very low, especially for large $\rho_{\d,\max}$. This indicates that, in \gls{HOMA}, the alarm message can be transmitted at high energy efficiency, at a cost of only a marginal backoff in the standard-traffic energy efficiency. 
        In our numerical optimization, the minimum required $(\EbNo)_\a$ is achieved when $\rho_\d= \rho_{\d,\max}$, and $\sf{P}_\a$ and $\sf{n}_\a$ are minimized such that $\max\{\epsilon_\amd, \epsilon_\afp\} \le 10^{-5}$. That is, one should let every user that detects the alarm event transmit at a low power using only few channel uses. 
		The reason is that, by having a large $K_\a$ (via increasing $\rho_\d$), one achieves a high effective \gls{SNR} $K_\a^2 \sf{P}_\a$, whereas for a fixed $K_\a$, one should minimize $\sf{P}_\a$ and $\sf{n}_\a$ to reduce the total energy $K_\a \sf{P}_\a \sf{n}_\a$. 
        This also explains why the required $(\EbNo)_\a$ decreases as $\sf{K}$ or $\rho_{\d,\max}$ increases, i.e., $K_\a$ increases. 

\begin{figure} [t]
			\centering
			\subfigure[Optimized alarm blocklength $\sf{n}_\a$]{
				\label{fig:sensitivity_opt_na}
				\begin{tikzpicture}
					\tikzstyle{every node}=[font=\footnotesize]
					\begin{axis}[%
						width=2.8in,
						height=1.8in,
						at={(0.759in,0.481in)},
						scale only axis,
						xmin=0,
						xmax=30000,
						xlabel style={font=\color{black},yshift=1ex},
						xlabel={\footnotesize Total number of users $\sf{K}$},
						ymin=0,
						ymax=2000,
						ylabel style={font=\color{black}, yshift=0ex},
						ylabel={\footnotesize Alarm blocklength $\sf{n}_\a^*$},
						axis background/.style={fill=white},
						xmajorgrids,
						ymajorgrids,
						legend style={at={(0.03,0.98)}, anchor=north west, row sep=-2.5pt, legend cell align=left, align=left, draw=white!15!black, nodes={scale=0.9},fill=white, fill opacity=0.2, text opacity=1, draw=none}
						]
						
						\addplot [color=blue,dashed,line width=1pt]
						table[row sep=crcr]{%
							1000 151 \\
							2500 152 \\
							5000 151 \\
							10000 151 \\
							15000 151 \\
							20000 151 \\
							25000 151 \\
							30000 148 \\
						};
						\node[rotate=0,color=blue] at (axis cs:.75e4,70) () {$\sf{n}_\a^*, \rho_{\d,\max} = 1$};
						
						\addplot [color=blue,dashdotted,line width=1pt]
						table[row sep=crcr]{%
							1000 526 \\
							2500 381 \\
							5000 335 \\
							10000 325 \\
							15000 325 \\
							20000 326 \\
							25000 321 \\
							30000 325 \\
						};
						\node[rotate=-4,color=blue] at (axis cs:.78e4,400) () {$\sf{n}_\a^*, \rho_{\d,\max} = 0.5$};
						
						\addplot [color=blue,dotted,line width=1pt]
						table[row sep=crcr]{%
							1000 2312 \\
							2500 1222 \\
							5000 734 \\
							10000 533 \\
							15000 456 \\
							20000 425 \\
							25000 407 \\
							30000 390 \\
						};
						\node[rotate=-7,color=blue] at (axis cs:1.6e4,530) () {$\sf{n}_\a^*, \rho_{\d,\max} = 0.1$};
						
						\addplot [color=blue,dotted,mark=square*,mark size=1.3,line width=1pt,mark options={solid,color=blue}]
						table[row sep=crcr]{%
							2500 18040 \\
							5000 14566 \\
							10000 2638 \\
							15000 1744 \\
							20000 944 \\
							25000 677 \\
							30000 521 \\
						};
						\node[rotate=-60,color=blue] at (axis cs:1.55e4,1450) () {$\sf{n}_\a^*, \rho_{\d,\max} = 0.01$};
						
						\addplot [color=red,line width=1pt]
						table[row sep=crcr]{%
							1000       22760 \\
							2500       19098 \\
							5000       14762 \\
							10000        7584 \\
							15000        1932 \\
							20000         964 \\
							25000         694 \\
							30000         526 \\
						};
						\node[rotate=-62,color=red] at (axis cs:1.75e4,1600) () {$\sf{n}_{\a,\max}$};
					\end{axis}
				\end{tikzpicture}%
			}
			\subfigure[Optimized power ratio $\sf{P}_\s/\sf{P}_\a$]{
				\label{fig:sensitivity_opt_Pa}
				\begin{tikzpicture}
					\tikzstyle{every node}=[font=\footnotesize]
					\begin{axis}[%
						width=2.9in,
						height=1.8in,
						at={(0.759in,0.481in)},
						scale only axis,
						xmin=0,
						xmax=30000,
						xlabel style={font=\color{black},yshift=1ex},
						xlabel={\footnotesize Total number of users $\sf{K}$},
						ymin=20,
						ymax=70,
						ytick={20, 30, 40, 50, 60, 70},
						ylabel style={font=\color{black}, yshift=-2ex},
						ylabel={\footnotesize $\sf{P}_\s/\sf{P}_\a$ (dB)},
						axis background/.style={fill=white},
						xmajorgrids,
						ymajorgrids,
						legend style={at={(0.99,0.26)}, anchor=south east, row sep=-2.5pt, legend cell align=left, align=left, draw=white!15!black, nodes={scale=0.9},fill=white, fill opacity=0.8, text opacity=1, draw=none}
						]
						\addplot [color=blue,dashed,line width=1pt]
						table[row sep=crcr]{%
							1000 40.073547916595942 \\ 
							2500 45.950764056423907 \\ 
							5000 51.838622072389114 \\ 
							10000 57.861556670885363 \\ 
							15000 61.632902367045688 \\ 
							20000 64.641855841001984 \\ 
							25000 67.070793004530515 \\ 
							30000 69.672530592081159 \\ 
						};
						\node[rotate=14,color=blue] at (axis cs:1.5e4,63.5) () {$\rho_{\d,\max} = 1$};
						
						\addplot [color=blue,dashdotted,line width=1pt]
						table[row sep=crcr]{%
							1000  38.634547109587402 \\
							2500 43.479607804787179 \\
							5000 48.995980290323018 \\
							10000 54.999788617109402 \\
							15000 58.657495048305918 \\
							20000 61.677126242100314 \\
							25000 63.948720211265652 \\
							30000 66.635806788604953 \\
						};
						\node[rotate=14,color=blue] at (axis cs:1.65e4,57) () {$\rho_{\d,\max} = 0.5$};
						
						\addplot [color=blue,dotted,line width=1pt]
						table[row sep=crcr]{%
							1000	27.100349828350545 \\
							2500 	25.018767593755626 \\
							5000	37.440053102649699 \\
							10000	42.648833609368126 \\
							15000	45.800511285066932 \\
							20000	48.595071903219718 \\
							25000	51.063694298033454 \\
							30000	53.259567422285897 \\
						};
						\node[rotate=13,color=blue] at (axis cs:1.75e4,44.5) () {$\rho_{\d,\max} = 0.1$};
						
						\addplot [color=blue,dotted,mark=square*,mark size=1.3,line width=1pt,mark options={solid,color=blue}]
						table[row sep=crcr]{%
							2500 	7.415631207889192 \\
							5000 22.476110273281904 \\
							10000 25.234895571834102 \\
							15000 28.557092834661432 \\
							20000 29.674341389276719 \\
							25000 31.258093978928002 \\
							30000 32.716311944812219 \\
						};
						\node[rotate=4.7,color=blue] at (axis cs:1.9e4,31.5) () {$\rho_{\d,\max} = 0.01$};
					\end{axis}
				\end{tikzpicture}%
			}
			\caption{The optimized alarm blocklength $\sf{n}_\a$ and power ratio $\sf{P}_\s/\sf{P}_\a$ for the setting in Fig.~\ref{fig:EbN0_sensitivity}.}
			\label{fig:sensitivity_opt_na_Pa}
		\end{figure}
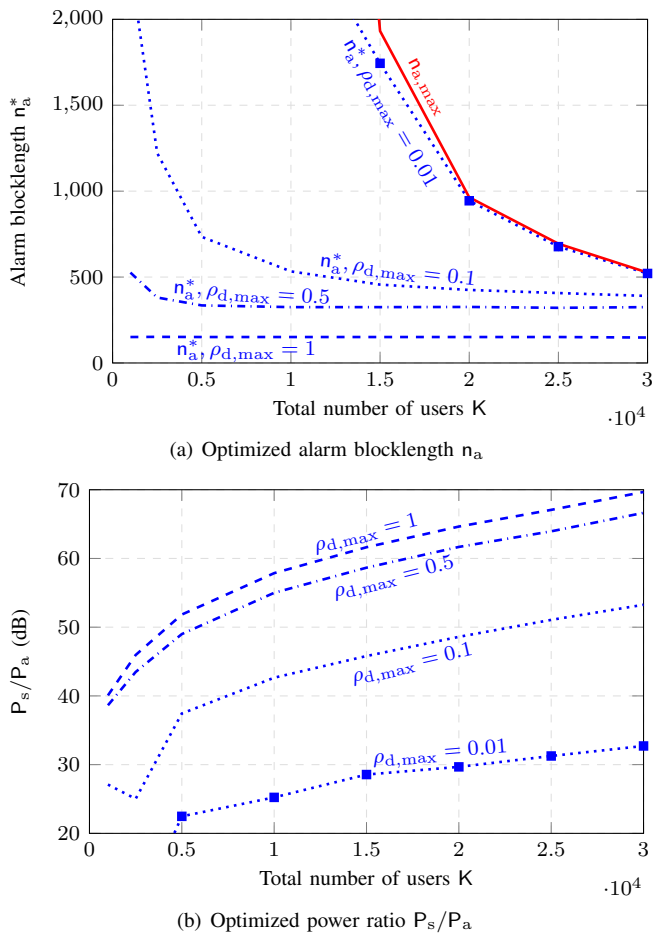

		In Fig.~\ref{fig:sensitivity_opt_na}, we show the numerically optimized alarm blocklength $\sf{n}_\a^*$ for the considered setting. We also plot $\sf{n}_{\a,\max}$ for reference. 
		If the users are not sensitive (i.e., $\rho_{\d,\max}$ is small), the active ones should transmit alarm codewords of length close to the maximum one, $\sf{n}_{\a,\max}$. As $\rho_{\d,\max}$ increases, alarm codewords of length progressively smaller than $\sf{n}_{\a,\max}$ suffice. 
		In Fig.~\ref{fig:sensitivity_opt_Pa}, we depict the ratio 
		$\sf{P}_\s/\sf{P}_\a$ for the optimized alarm power~$\sf{P}_\a$. 
		We observe that the optimized $\sf{P}_\a$ is much smaller than $\sf{P}_\s$, and the ratio $\sf{P}_\s/\sf{P}_\a$ increases as $\sf{K}$ or $\rho_{\d,\max}$ increases. 
        \revise{We further observe from numerical evaluation that if $\rho_{\d,\max}$ is high, i.e., many users transmit the alarm message, the bottleneck is to satisfy $P_\amd \le 10^{-5}$. However, the bottleneck becomes the AFP requirement $P_\afp \le 10^{-5}$ if $\rho_{\d,\max}$ is low, i.e., few users transmit the alarm message.}
		
		\subsubsection{Impact of the Dynamic Range}
\begin{figure}[t]
			\centering
			\begin{tikzpicture}
				\tikzstyle{every node}=[font=\footnotesize]
				\begin{axis}[%
					width=2.9in,
					height=1.8in,
					at={(0.759in,0.481in)},
					scale only axis,
					xmin=0,
					xmax=30000,
					xlabel style={font=\color{black},yshift=1ex},
					xlabel={\footnotesize Total number of users $\sf{K}$},
					ymin=-30,
					ymax=10,
					ylabel style={font=\color{black}, yshift=-2ex},
					ylabel={\footnotesize Required $E_{\rm b}/N_0$ (dB)},
					axis background/.style={fill=white},
					xmajorgrids,
					ymajorgrids,
					legend style={at={(0.01,0.01)}, anchor=south west, row sep=-2.5pt, legend cell align=left, align=left, draw=white!15!black, nodes={scale=0.95},fill=white, fill opacity=0.8, text opacity=1, draw=none}
					]
					\addplot [color=red,line width=1pt]
					table[row sep=crcr]{
						1000 1.638818835868710 \\ 
						2500 1.093003773659495 \\ 
						5000 1.046901113925378 \\ 
						10000 1.108300355924746 \\ 
						15000 1.241000190960597 \\ 
						20000 1.750954301244553 \\ 
						25000 2.419138257681345 \\ 
						30000 3.174586008254130 \\ 
					};
					\node[rotate=.5,color=red] at (axis cs:.83e4,2.5) () {\scriptsize $(\EbNo)_{\s}^* + 0.1$~dB};
					
					\addplot [color=blue,dashdotted,line width=1pt]
					table[row sep=crcr]{%
						1000 -14.622982195687996 \\
						2500  -18.602333375364935 \\
						5000 -21.612451758486770 \\
						10000 -24.622988972511735 \\
						15000 -26.383838627586272 \\
						20000 -27.632562709228864 \\
						25000 -28.601399885797377 \\
						30000  -29.392620668723794 \\
					};
					\node[rotate=-5,color=blue] at (axis cs:2.5e4,-27) () {$(\EbNo)_{\a}, \psi = \infty$};
					
					\addplot [color=blue,dashdotted,line width=1pt,mark=*,mark size=2,mark color=blue,mark options={solid}]
					table[row sep=crcr]{%
						1000 -12.8475 \\ 
						2500  -12.7184  \\
						5000 -12.0902 \\ 
						10000 -11.974996918805061 \\
						15000 -11.878183568399493 \\
						20000 -11.598812286407565 \\
						25000 -11.251962892010908 \\
						30000  -10.848511927163827 \\
					};
					\node[rotate=2.5,color=blue] at (axis cs:2.4e4,-13.5) () {$(\EbNo)_{\a}, \psi = 30$~dB};
					
					\addplot [color=blue,dashdotted,line width=1pt,mark=triangle,mark size=2,mark color=blue,mark options={solid}]
					table[row sep=crcr]{%
						1000 -14.622982195687996 \\
						2500  -18.602333375364935 \\
						5000 -21.4916 \\ 
						10000 -22.9884 \\
						15000 -22.92 \\
						20000 -22.7377 \\
						25000 -22.0038 \\
						30000  -21.0809 \\ 
					};
					\node[rotate=2.5,color=blue] at (axis cs:2.4e4,-20.5) () {$(\EbNo)_{\a}, \psi = 50$~dB};

					\addplot [color=blue,dashdotted,line width=1pt,mark=diamond,mark size=2.5,mark color=blue,mark options={solid}]
					table[row sep=crcr]{%
						1000 -1.139827156973354 \\
						2500  -1.311926184165543 \\
						5000 -1.271712036805923 \\
						10000 -1.196197587158053 \\
						15000 -1.119092980534504 \\
						20000 -0.824060962915155 \\
						25000 -0.457678681752725 \\
						30000  -0.013791194766611 \\
					};
					\node[rotate=2.5,color=blue] at (axis cs:2.4e4,-2.2) () {$(\EbNo)_{\a}, \psi = 10$~dB};
					
					\addplot [color=blue,dashdotted,line width=1pt,mark=square,mark size=2,mark color=blue,mark options={solid}]
					table[row sep=crcr]{%
						1000 4.810128568979280 \\
						2500  4.578782302094908 \\
						5000 4.583249584020734 \\
						10000 4.625529927710488 \\
						15000 4.714285219386568 \\
						20000 5.056919340922491 \\
						25000 5.460007321010096 \\
						30000  5.924321485368750 \\
					};
					\node[rotate=1.5,color=blue] at (axis cs:2.37e4,7.3) () {$(\EbNo)_{\a}, \psi = 0$~dB};
				\end{axis}
			\end{tikzpicture}%
			\caption{The minimum $(\EbNo)_\a$ required for \gls{HOMA} to satisfy $\max\{\epsilon_\amd, \epsilon_\afp\} \le 10^{-5}$ when $(\EbNo)_{\s} = (\EbNo)_{\s}^* + 0.1$~dB for different dynamic range $\psi$. Here, $\sf{M}_\a = 2^3$, $\sf{M}_\s = 2^{100}$, $\rho_\s = 0.01$, and $\rho_{\d,\max} = 1$.} 
			\label{fig:EbN0_dynamic}
		\end{figure}
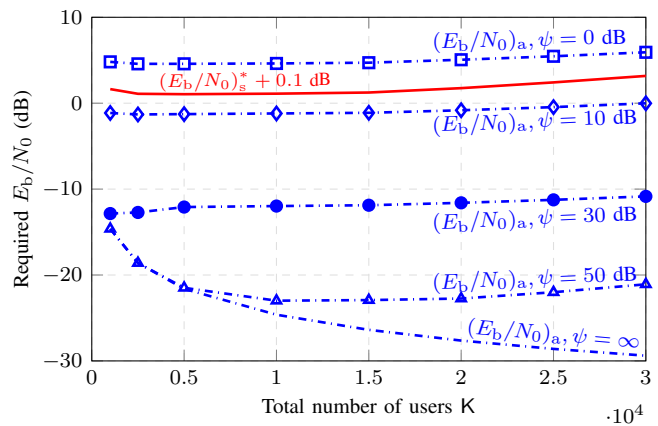

		The power difference between the alarm and the standard blocks requires the user devices to support a wide dynamic range. 
		Recently designed transmitters for narrow-band \gls{IoT} applications have the dynamic range from $34.1$~dB~\cite{Jin2017efficient} 
		to $72.3$~dB~\cite{Mahalingam2022large}. To account for the limited dynamic range of IoT devices, we impose the additional constraint $\sf{P}_\s/\sf{P}_\a \le \psi$ in the minimization~\revise{\eqref{eq:minEbNoa}} of the required $(\EbNo)_\a$. In Fig.~\ref{fig:EbN0_dynamic}, we plot the minimum required $(\EbNo)_\a$ for $\rho_{\d,\max} = 1$ and $\psi \in \{0,10,30,50,\infty\}$~dB. The case of infinite dynamic range corresponds to the setting in Fig.~\ref{fig:EbN0_sensitivity}. We see that a narrower dynamic range leads to a higher required $(\EbNo)_\a$. If the users transmit at equal power over the two blocks, i.e., $\psi = 0$~dB, the alarm traffic requires a higher energy per bit than the standard traffic. 
		
\begin{figure} [t]
			\centering
			\subfigure[Optimized alarm blocklength $\sf{n}_\a$]{
				\label{fig:dynamic_opt_na}
				\begin{tikzpicture}
					\tikzstyle{every node}=[font=\footnotesize]
					\begin{groupplot}[
						group style={
							group name=my fancy plots,
							group size=1 by 2,
							xticklabels at=edge bottom,
							vertical sep=0pt
						},
						width=2.8in,
						height=1.8in,
						at={(0.759in,0.481in)},
						scale only axis,
						xmin=0,
						xmax=30000,
						ymin=10,
						ymax=400,
						axis background/.style={fill=white},
						xmajorgrids,
						ymajorgrids,
						xtick align=inside,
						legend style={at={(0.98,0.98)}, anchor=north east, row sep=-2.5pt, legend cell align=left, align=left, draw=white!15!black, nodes={scale=0.9},fill=white, fill opacity=0.8, text opacity=1, draw=none}
						]
						
						\nextgroupplot[ymin=140,ymax=160,
						ytick={150,160},
						axis x line=top, 
						xtick align=inside,
						axis y discontinuity=parallel,
						height=.5in]
						\addplot [color=blue,dashdotted,line width=1pt]
						table[row sep=crcr]{%
							1000 151 \\
							2500 152 \\
							5000 151 \\
							10000 151 \\
							15000 151 \\
							20000 151 \\
							25000 151 \\
							30000 148 \\
						};
						
						\node[rotate=0,color=black] at (axis cs:2e4,154) () {$\psi = \infty$~dB};
						
						
						\nextgroupplot[ymin=10,ymax=41,
						ytick={10,20,30,40},
						axis x line=bottom, 
						xtick align=inside,
						height=1.1in,
						xlabel style={font=\color{black},yshift=1ex},
						xlabel={\footnotesize Total number of users $\sf{K}$},
						ylabel style={font=\color{black}, yshift=-1.5ex, xshift=3ex},
						ylabel={\footnotesize Alarm blocklength $\sf{n}_\a$},
						]
						\addplot [color=blue,dashdotted,line width=1pt,mark=triangle,mark size=2,mark color=blue,mark options={solid}]
						table[row sep=crcr]{%
							1000 151 \\
							2500  152 \\
							5000 100 \\
							10000 35 \\
							15000 22 \\
							20000 23 \\
							25000 24 \\
							30000  23 \\
						};
						
						\addplot [color=blue,dashdotted,line width=1pt,mark=*,mark size=2,mark color=blue,mark options={solid}]
						table[row sep=crcr]{%
							1000 32 \\
							2500  15 \\
							5000 19 \\
							10000 21 \\
							15000 21 \\
							20000 21 \\
							25000 21 \\
							30000  21 \\
						};
						
						
						\addplot [color=blue,dashdotted,line width=1pt,mark=diamond,mark size=2.5,mark color=blue,mark options={solid}]
						table[row sep=crcr]{%
							1000 14 \\
							2500  15 \\
							5000 14 \\
							10000 15 \\
							15000 15 \\
							20000 13 \\
							25000 13 \\
							30000  14 \\
						};
						
						\addplot [color=blue,dashdotted,line width=1pt,mark=square,mark size=2,mark color=blue,mark options={solid}]
						table[row sep=crcr]{%
							1000 11 \\
							2500  11 \\
							5000 12 \\
							10000 11 \\
							15000 11 \\
							20000 11 \\
							25000 11 \\
							30000  12 \\
						};
						
						\draw[-latex] (axis cs:1.25e4,10.2) -- node[pos=1,right] () {$\psi = 0,10,30,50$~dB} (axis cs:1.25e4,35);
					\end{groupplot}
				\end{tikzpicture}%
			}
			\subfigure[Optimized average number of alarm users $\rho_\d \sf{K}$]{
				\label{fig:dynamic_opt_Pa}
				\begin{tikzpicture}
					\tikzstyle{every node}=[font=\footnotesize]
					\begin{axis}[%
						width=2.8in,
						height=1.8in,
						at={(0.759in,0.481in)},
						scale only axis,
						xmin=0,
						xmax=30000,
						xlabel style={font=\color{black},yshift=1ex},
						xlabel={\footnotesize Total number of users $\sf{K}$},
						ymin=100,
						ymax=40000,
						ytick={1e2, 1e3, 1e4, 4e4},
                        yticklabels={$10^2$,$10^3$,$10^4$,$4\cdot 10^4$},
						ymode=log,
						ylabel style={font=\color{black}, yshift=-2ex,xshift=-.3cm},
						ylabel={\footnotesize Average number of alarm users $\rho_\d \sf{K}$},
						axis background/.style={fill=white},
						xmajorgrids,
						ymajorgrids,
						legend style={at={(0.99,0.99)}, anchor=north east, row sep=-2.5pt, legend cell align=left, align=left, draw=white!15!black, nodes={scale=0.9},fill=white, fill opacity=0.8, text opacity=1, draw=none}
						]
						
						\addplot [color=blue,dashdotted,line width=1pt]
						table[row sep=crcr]{%
							1000         1000 \\ 
							2500        2500 \\ 
							5000        5000 \\ 
							10000        10000 \\ 
							15000        15000 \\ 
							20000       20000 \\
							25000       25000 \\
							30000       30000 \\
						};
						
						\addplot [color=blue,dashdotted,line width=1pt,mark=*,mark size=2,mark color=blue,mark options={solid}]
						table[row sep=crcr]{%
							1000 1000 \\ 
							2500 2493.425 \\ 
							5000 2.3058e+03 \\ 
							10000 2105.665466516580 \\
							15000 2088.337687828890 \\
							20000 1980.352237315700 \\
							25000 1935.460711957025 \\
							30000 1874.579980288290 \\
						};
						
						\addplot [color=blue,dashdotted,line width=1pt,mark=triangle,mark size=2,mark color=blue,mark options={solid}]
						table[row sep=crcr]{%
							1000 1000 \\ 
							2500 2498 \\ 
							5000 5000 \\ 
							10000 10000 \\
							15000 15000 \\
							20000 13913 \\
							25000 13542 \\
							30000 14901 \\
						};
						
						
						\addplot [color=blue,dashdotted,line width=1pt,mark=diamond,mark size=2.5,mark color=blue,mark options={solid}]
						table[row sep=crcr]{%
							1000 338.8809503964170 \\
							2500 344.6998171689025 \\
							5000 376.7475533473450 \\
							10000 352.7639968275500 \\
							15000 348.2769493375350 \\
							20000 331.4610657551400 \\
							25000 356.8036004513750 \\
							30000 308.3713594734000 \\
						};
						
						\addplot [color=blue,dashdotted,line width=1pt,mark=square,mark size=2,mark color=blue,mark options={solid}]
						table[row sep=crcr]{%
							1000 169.7545904168820 \\
							2500 182.5023271277325 \\
							5000 169.2475372290000 \\
							10000 183.8287825326800 \\
							15000 181.9780707873450 \\
							20000 175.1003675634800 \\
							25000 164.7318061357250 \\
							30000 141.2085711543300 \\
						};
						
						\draw[-latex] (axis cs:2.3e4,100) -- node[pos=.65,left] () {$\psi = 0,10,30,50,\infty$~dB} (axis cs:2.3e4,32000);
					\end{axis}
				\end{tikzpicture}%
			}
				\caption{The optimized alarm blocklength $\sf{n}_\a$ and average number of alarm users $\rho_\d \sf{K}$ for the setting in Fig.~\ref{fig:EbN0_dynamic}.}
				\label{fig:dynamic_opt_na_Pa}
			\end{figure}
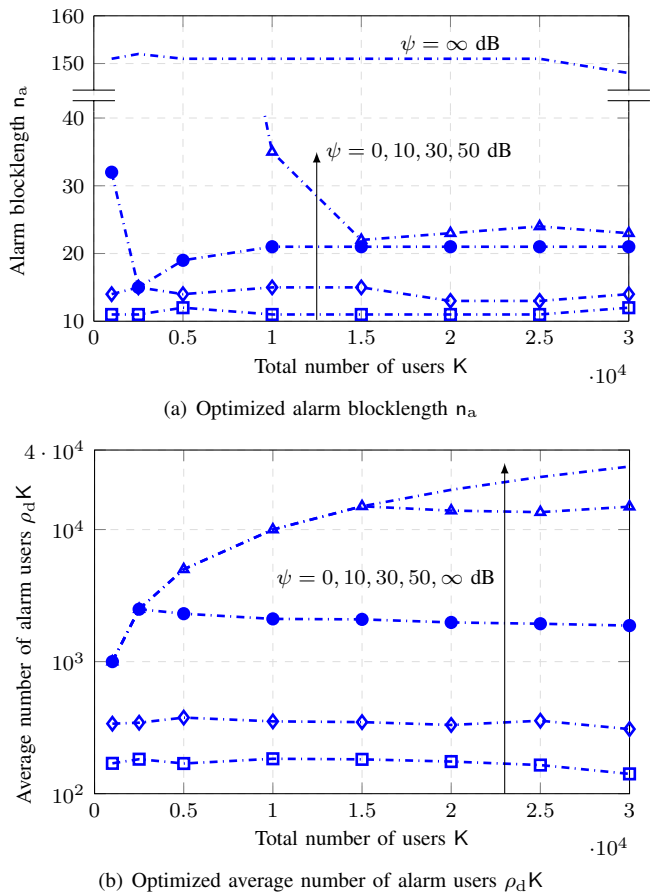

		We observe that for each $\sf{K}$ and $\sf{n}_\a$, if $\psi$ is large enough such that the power ratios depicted in Fig.~\ref{fig:sensitivity_opt_Pa} are achievable, one should let $\rho_\d = 1$ and use the optimized $\sf{P}_\a$ for the $\psi = \infty$ case. Otherwise, the required $(\EbNo)_\a$ is minimized \revise{with $\sf{P}_\a$ equal to its minimum value $\sf{P}_\s/ \psi$ and $\rho_\d$ equal to the smallest value} such that $\max\{\epsilon_\amd, \epsilon_\afp\} \le 10^{-5}$. In Fig.~\ref{fig:dynamic_opt_na_Pa}, we show the optimized alarm blocklength $\sf{n}_\a$ and optimized average number of alarm users $\rho_\d \sf{K}$ obtained from our numerical optimization. As shown in Fig.~\ref{fig:dynamic_opt_na}, when the dynamic range is limited to $50$~dB and $\sf{K} \ge 10^4$, the optimized alarm blocklength is significantly lower than for the case of infinite dynamic range. This is because when each alarm user transmits at a higher power, the codeword can be shortened. However, as shown in Fig.~\ref{fig:EbN0_dynamic}, this results in higher required $(\EbNo)_\a$. Fig.~\ref{fig:dynamic_opt_Pa} shows that as we increase $\psi$, the average number of alarm users increases. Indeed, a lower $\psi$ leads to a smaller optimized $\sf{n}_\a$ and $\rho_\d$. 
			
			\subsection{\gls{HNOMA}}
			\label{sec:H-NOMA}
			For \gls{HNOMA}, we set $\sf{P}_\s$ such that $(\EbNo)_\s = (\EbNo)_\s^* + \delta$~dB and find the minimum $\rho_\d$ and $\sf{P}_\a$ required to satisfy the requirements of both traffic types. We choose $k_{\s,\ell}$ and $k_{\s,u}$ such that $1 - \sum_{k = k_{\s,\ell}}^{k_{\s,u}}{\rm Bino}(k;\sf{K},\rho_\s) < 10^{-10}$, and choose $\sf{P}'_\s$ for each $\sf{P}_\s$ such that $\frac{\Gamma(\sf{n}/2,\sf{n} \sf{P}_\s/(2\sf{P}'_\s))}{\Gamma(\sf{n}/2)} < 10^{-5}$. It turns out that the main challenge is to satisfy the target SMD and SFP probabilities when there is an alarm. Indeed, although the alarm message $W_\a$ can usually be reliably decoded, the number of alarm users $K_\a$ is estimated incorrectly with significant probability. The probability of wrongly estimating the number of alarm users given no AMD can be computed as $\P{\argmin_{k} \|\vec Y - k \revise{\vec C_{\a,W_\a}}\|^2 \ne K_\a}$. Note that this probability also accounts for a reestimation of $K_\a$ after correctly decoding $W_\a$. With $K_\s = 100$ and $\sf{P}_\a = \sf{P}_\s/\psi$, we have that $\P{\argmin_{k} \|\vec Y - k \revise{\vec C_{\a,W_\a}}\|^2 \ne K_\a}$ is equal to $0.276$ for $\psi = 20$~dB and $0.426$ for $\psi = 30$~dB. With $K_\s = 200$, this probability is $0.306$ for $\psi = 20$~dB and $0.433$ for $\psi = 30$~dB. (In our bound, this is reflected by the fact that $\alpha(k_\a,\hat{k}_\a,k_\s)$ is significantly large for $k_\a \ne \hat{k}_\a$, especially if $k_\s$ is large.) 
			This leads to a high residual interference $K_\a \revise{\vec C_{\a,W_\a}} \!-\! \widehat{K}_\a \revise{\vec C_{\a,\widehat{W}_{\a}}}$ when decoding the standard messages. \revise{We note that lowering $P_\amd$ and $P_\afp$ does not resolve this issue because, as we have seen, $K_\a$ is estimated incorrectly with significant probability even if ${\widehat{W}_{\a}} \!=\! {{W}_\a}$.} 

			With the same $(\EbNo)_\s$ backoff $\delta = 0.1$~dB as considered for \gls{HOMA}, \gls{HNOMA} cannot satisfy the reliability requirements for both traffic types unless the estimation of $K_\a$ is reliable, which occurs when $\rho_\d = 1$, i.e., $K_\a=\sf{K}$, or when $\sf{P}_\a$ is high, i.e., comparable to $\sf{P}_\s$.  
			In Fig.~\ref{fig:EbN0_HNOMA}, we plot the minimum required $(\EbNo)_\a$ for \gls{HNOMA} in these two cases. Specifically, we consider $\rho_\d = 1$ and minimize $\sf{P}_\a$, and also consider $\sf{P}_\a = \sf{P}_\s$ and minimize $\rho_\d$. As a comparison, we depict the corresponding $(\EbNo)_\a$ values for \gls{HOMA}, which are obtained by considering the two cases $\psi = \infty$ and $\psi = 0$~dB. 
			For $\sf{P}_\a = \sf{P}_\s$, the $(\EbNo)_\a$ of \gls{HNOMA} is lower-bounded by
			$(\EbNo)_\a = \frac{\sf{n} \sf{P}_\a \rho_\d \sf{K}}{2 \log_2 \sf{M}_\a} \ge \frac{\sf{n} \sf{P}_\s (1 - \epsilon_{\amd}^{1/\sf{K}}) \sf{K}}{2 \log_2 \sf{M}_\a} \revise{\ge \frac{\sf{n} \sf{P}_\s (1 - (10^{-5})^{1/\sf{K}}) \sf{K}}{2 \log_2 \sf{M}_\a}}$ since $\rho_\d \ge 1 - \epsilon_{\amd}^{1/\sf{K}}$ (see Remark~\ref{rem:rho_d}). 
            The minimum required $(\EbNo)_\a$ of \gls{HNOMA} is slightly higher than this lower bound, and is much higher than the minimum required $(\EbNo)_\a$ of \gls{HOMA}. For $\rho_\d = 1$, the minimum required $(\EbNo)_\a$ of \gls{HNOMA} is also significantly higher than \gls{HOMA}. In this case, the optimized $\sf{P}_\a$ is similar in both schemes, but the alarm codeword length $\sf{n}_\a$ in \gls{HOMA} is much shorter than that in \gls{HNOMA} (i.e., $\sf{n}$). This leads to a significant difference in energy efficiency. 
			\begin{figure}[t]
				\centering
				\begin{tikzpicture}
					\tikzstyle{every node}=[font=\footnotesize]
					\begin{axis}[%
						width=2.9in,
						height=1.8in,
						at={(0.759in,0.481in)},
						scale only axis,
						xmin=0,
						xmax=30000,
						xlabel style={font=\color{black},yshift=1ex},
						xlabel={\footnotesize Total number of users $\sf{K}$},
						ymin=-30,
						ymax=34,
						ytick={-30, -20, -10, 0, 10, 20, 30},
						ylabel style={font=\color{black}, yshift=-2ex},
						ylabel={\footnotesize Required $E_{\rm b}/N_0$ (dB)},
						axis background/.style={fill=white},
						xmajorgrids,
						ymajorgrids,
						legend style={at={(0.01,0.01)}, anchor=south west, row sep=-2.5pt, legend cell align=left, align=left, draw=white!15!black, nodes={scale=0.95},fill=white, fill opacity=0.8, text opacity=1, draw=none}
						]
						\addplot [color=red,line width=1pt]
						table[row sep=crcr]{
							1000 1.638818835868710 \\ 
							2500 1.093003773659495 \\ 
							5000 1.046901113925378 \\ 
							10000 1.108300355924746 \\ 
							15000 1.241000190960597 \\ 
							20000 1.750954301244553 \\ 
							25000 2.419138257681345 \\ 
							30000 3.174586008254130 \\ 
						};
						\node[rotate=.5,color=red] at (axis cs:.83e4,-2) () {\scriptsize $(\EbNo)_{\s}^* + 0.1$~dB};
						
						\addplot [color=blue,dashdotted,line width=1pt]
						table[row sep=crcr]{%
							1000 -14.622982195687996 \\
							2500  -18.602333375364935 \\
							5000 -21.612451758486770 \\
							10000 -24.622988972511735 \\
							15000 -26.383838627586272 \\
							20000 -27.632562709228864 \\
							25000 -28.601399885797377 \\
							30000  -29.392620668723794 \\
						};
						\node[rotate=-4,color=blue] at (axis cs:2.2e4,-25.5) () {$(\EbNo)_{\a}, \rho_\d = 1$, \gls{HOMA}};
						
						\addplot [color=black,dashdotted,line width=1pt]
						table[row sep=crcr]{%
							1000 -6.8509 \\
							2500  -10.3906 \\
							5000 -12.7134 \\
							10000 -14.5805 \\
							15000 -15.3806 \\
							20000 -15.5555 \\
							25000 -15.6822 \\
							30000  -15.9856 \\
						};
						\node[rotate=-1,color=black] at (axis cs:2.2e4,-13) () {$(\EbNo)_{\a}, \rho_\d = 1$, \gls{HNOMA}};
						
						\addplot [color=black,dashed,line width=1pt]
						table[row sep=crcr]{%
							1000 27.6195 \\
							2500 27.913723950527071 \\
							5000 27.885925279620153 \\
							10000  28.0009  \\
							15000 28.1810  \\
							20000 28.729546428407161 \\ 
							25000 29.439075674935566 \\ 
							30000 30.236383955352096 \\ 
						};
						\node[rotate=2,color=black] at (axis cs:1.8e4,24.5) () {$(\EbNo)_{\a}$ lower bound, $\sf{P}_\a = \sf{P}_\s$, \gls{HNOMA}};
						
						\addplot [color=black,line width=1pt]
						table[row sep=crcr]{%
							1000 27.3163 \\
							2500 26.7886 \\
							5000 26.7489 \\
							10000 26.8183 \\
							15000 26.9534 \\
							20000 27.4617 \\
							25000 28.4331 \\
							30000 28.8914 \\
						};
						\node[rotate=.5,color=black] at (axis cs:1e4,30.5) () {$(\EbNo)_{\a}, \sf{P}_\a = \sf{P}_\s$, \gls{HNOMA}};
						
						\addplot [color=blue,dashdotted,line width=1pt,mark=square,mark size=2,mark color=blue,mark options={solid}]
						table[row sep=crcr]{%
							1000 4.810128568979280 \\
							2500  4.578782302094908 \\
							5000 4.583249584020734 \\
							10000 4.625529927710488 \\
							15000 4.714285219386568 \\	
							20000 5.056919340922491 \\
							25000 5.460007321010096 \\
							30000  5.924321485368750 \\
						};
						\node[rotate=1,color=blue] at (axis cs:2.2e4,8.5) () {$(\EbNo)_{\a}, \sf{P}_\a = \sf{P}_\s$, \gls{HOMA}};
					\end{axis}
				\end{tikzpicture}%
				\caption{The minimum $(\EbNo)_\a$ required for~\gls{HOMA} and \gls{HNOMA} to satisfy $\max\{\revise{\epsilon_{\smd | \Ac}, \epsilon_{\smd | \bar{\Ac}}, \epsilon_{\sfp,\Ac}, \epsilon_{\sfp,\bar{\Ac}}}\} \le 10^{-1}$ and $\max\{\epsilon_\amd, \epsilon_\afp\} \le 10^{-5}$ when $(\EbNo)_{\s} = (\EbNo)_{\s}^* + 0.1$~dB for two cases: i) $\rho_\d = 1$ and optimized $\sf{P}_\a$, and ii) $\sf{P}_\a = \sf{P}_\s$ and optimized $\rho_\d$. 
					Here, $\sf{M}_\a = 2^3$, $\sf{M}_\s = 2^{100}$, $\rho_\s = 0.01$, and $\rho_{\d,\max} = 1$.} 
				\label{fig:EbN0_HNOMA}
			\end{figure}
			
			Our results suggest that reliability diversity~\cite{Popovski2018} is hard to exploit in nonorthogonal network slicing between massive and critical~\gls{IoT} over the Gaussian MAC. This supports the claim in~\cite{Popovski2018} that nonorthogonal network slicing between \gls{URLLC} and \gls{mMTC} may be problematic. While~\cite{Popovski2018} predicted that the main challenge is to guarantee reliability for \gls{URLLC} devices, we point out that reliable message decoding for \gls{URLLC} devices is not enough to effectively perform interference cancellation. In our setup, the number of active \gls{URLLC} devices (i.e., the number of alarm users) also needs to be estimated reliably.\footnote{\gls{HOMA} also has advantage in terms of latency for the alarm traffic since the alarm codeword occupies only a small fraction of a frame and it is decoded first.} 

            \revise{We further note that interference cancellation is not strictly needed for \gls{HNOMA}. Without interference cancellation, the receiver decodes the list of standard messages by treating the alarm signal as noise. Our bound in Theorem~\ref{th:HNOMA} can be easily adapted to this case. However, due to the interference from the alarm signal, the standard-traffic reliability requirements are difficult to achieve with a small backoff. For the setting in Fig.~\ref{fig:EbN0_HNOMA} with $\rho_\d = 1$, 
            the backoff has to be progressively increased from $0.18$ to $0.89$~dB as $\sf{K}$ grows from $1000$ to $30000$.}
			\section{Conclusions and Future Works}\label{sec:conclusion}
			We investigated massive and critical \gls{IoT} in a setting where both standard \gls{UMA} traffic and alarm traffic are present. 
			We considered a random and unknown number of active users and 
			accounted for misdetections and false positives in both traffic types. For the Gaussian \gls{MAC}, our results show that both traffic types can coexist with high energy efficiency by means of orthogonal network slicing, provided that a large number of users transmit the alarm message, and that the transmit power of the alarm message is much smaller than that of the standard message. On the contrary, nonorthogonal network slicing is energy inefficient due to the residual interference from the alarm signal when decoding the standard messages, caused by unreliable estimation of the number of alarm users. These results indicate that it is hard to exploit reliability diversity to perform nonorthogonal network slicing between massive and critical \gls{IoT}, and that orthogonal network slicing is \revise{preferable}.

            Our conclusions pertain to the Gaussian \gls{MAC} and to the considered random-coding scheme. In more general/practical settings, \gls{HNOMA} might have advantages over \gls{HOMA}. First, in this paper, we aim to satisfy the SMD and SFP requirements in both alarm and no alarm states, and the bottleneck of \gls{HNOMA} is the alarm state. If one considers average SMD and SFP probabilities over the two states, 
            \gls{HNOMA} can still satisfy the requirements if the alarm event is rare. Second, in our setting, the detection of the alarm message benefits from the coherent addition of the common alarm codeword transmitted by many users, and thus only few channel uses suffice for the alarm traffic in \gls{HOMA}. In practice, the channel may cause a mismatch, such as a phase rotation \revise{or a scaling factor}, between the signal transmitted from different users. Furthermore, multiple alarm messages with different reliability guarantees may need to be simultaneously reported. These aspects \revise{need to be} be taken into account.
            Third, one can consider dynamic power adaptation based on the presence/absence of alarm messages. For example, one can let the users increase the transmit power of the standard message if the alarm message is not present. 

			\appendices
			\section{Proof of Theorem~\ref{th:alarm_block}} \label{proof:HOMA}
   
			The following well-known results will be used in our proof.
			
			\begin{lemma}[{Change of measure~\cite[Lemma~4]{Ohnishi2021}}] \label{lem:change_measure}
				Let $p$ and $q$ be two probability measures. Consider a random variable $X$ supported on $\mathcal{H}$ and a function $f \colon \mathcal{H} \to [0,1]$. It holds that $\E[p]{f(X)} \le \E[q]{f(X)} + d_{\rm TV}(p,q)$, where $d_{\rm TV}(p,q)$ is the total variation distance~\cite[Sec.~2]{Gibbs2002_on_choosing} between $p$ and~$q$.
			\end{lemma}

			\begin{lemma}[{Chernoff bound~\cite[Th. 6.2.7]{DeGroot2012ProbStats}}] \label{lem:Chernoff}
				For a random variable $X$ with moment-generating function $\E{e^{t X}}$ defined for all $|t| \le b$, it holds for all $\lambda \in [0,b]$ that $\P{X \le x} \le e^{\lambda x} \E{e^{-\lambda X}}$.
			\end{lemma}
			
			\begin{lemma} \label{lem:chi2}
				Let $\vec X  \sim \mathcal{N}(\vec \mu,\sigma^2\vec I_n)$. It holds that 
				\begin{equation} \label{eq:chi2_MGF}
					\E{e^{-\gamma \|\vec X\|^2}} = (1+2\gamma\sigma^2)^{-n/2} \exp\Big(-\frac{\gamma\|\vec \mu\|^2}{1+2\gamma\sigma^2}\Big) 
				\end{equation} 
                for every $\gamma > -\frac{1}{2\sigma^2}$.
				Furthermore, if $\vec \mu = \vec 0$, it holds that
				\begin{equation} \label{eq:chi2_CCDF}
					\P{\|\vec X\|^2 > y} = \frac{\Gamma(n/2,y/(2\sigma^2))}{\Gamma(n/2)}.
				\end{equation}
			\end{lemma}
			\begin{IEEEproof}
				Note that $\norm{\vec X}/\sigma$ follows a noncentral chi-squared distribution with $n$ degrees of freedom and noncentrality parameter $\norm{\vec \mu}/\sigma^2$.  The results in~\eqref{eq:chi2_MGF} and~\eqref{eq:chi2_CCDF} follows straightforwardly from the expressions of moment generating function and complementary cumulative distribution function, respectively, of this distribution.
			\end{IEEEproof}

			\subsection{Proof of the Bound~\eqref{eq:RCU_AMD} on $P_\amd$}
			\label{proof:OMA_AMD}
			The AMD probability averaged over the Gaussian codebook ensemble is computed as
			\begin{equation}
				P_\amd = \EE_{K_\a, \Cc_\a}\big[\mathbbm{1}\big\{ \widehat{W}_{\a} \ne W_\a\big\} \big \vert \Ac\big]. \label{eq:tmp13}
			\end{equation}
			
			\subsubsection{A Change of Measure}
   
            We first perform a change of measure as in~\cite{Polyanskiy2017,Ngo2022}. Specifically, we replace the measure over which the expectation in~\eqref{eq:tmp13} is taken by the one under which 
			$\vec X_\a=\vec C_{\a, W_\a}$ instead of $\vec X_\a=\vec C_{\a, W_\a} \ind{\norm{\vec C_{\a, W_\a}} \le \sf{n}_\a \sf{P}_\a}$. Furthermore, under the new measure, there are at least $k_{\a,\ell}$ and at most $k_{\a,u}$ users transmitting the alarm message, i.e., $K_{\a} \in [k_{\a,\ell} : k_{\a,u}]$. It then follows from~\cite[Eq.~(41)]{Kowshik2020fundamental} that the total variation between the original measure and the new one is upper-bounded by $\nu_0 \triangleq \P{\norm{\vec C_{\a, W_\a}}> \sf{n}_{\a}\sf{P}_\a} + \P{K_\a \ne [k_{\a,\ell} : k_{\a,u}]}$. Since $\vec C_{\a, W_\a} \sim \mathcal{N}(\mathbf{0},\sf{P}'_\a\vec I_{\sf{n}_\a})$, it follows from~\eqref{eq:chi2_CCDF} that $\P{\norm{\vec C_{\a, W_\a}}> \sf{n}_{\a}\sf{P}_\a} = \frac{\Gamma(\sf{n}_\a/2,\sf{n}_\a \sf{P}_\a/(2\sf{P}'_\a))}{\Gamma(\sf{n}_\a/2)}$.  Furthermore,  $\P{K_\a \ne [k_{\a,\ell} : k_{\a,u}]} = 1 - \sum_{k = k_{\a,\ell}}^{k_{\a,u}}P_{K_\a}(k)$, where $P_{K_\a}(k) = {\rm Bino}(k;\sf{K},\rho_\d)$. Therefore, $\nu_0$ is given by~\eqref{eq:p0}. By applying Lemma~\ref{lem:change_measure} to the random quantity $\mathbbm{1}\big\{ \widehat{W}_{\a} \!\ne\! W_\a\big\}$, we consider implicitly the new measure from now on at a cost of adding $\nu_0$ to the original expectation in~\eqref{eq:tmp13}. That is
			\begin{align} \label{eq:tmp24}
				P_\amd &= \sum_{k_\a = k_{\a,\ell}}^{k_{\a,u}} P_{K_\a}(k_\a) \E[\Cc_\a]{ \ind{\widehat{W}_{\a} \ne W_\a} \big\vert \Ac, K_\a = k_\a} \notag \\
                &\quad+ \nu_0 
			\end{align}
			where 
			the expectation 
            is taken under the assumption that $\vec X_\a = \vec C_{\a, W_\a}$. 
			
			\subsubsection{Expanding $\EE_{\Cc_\a}\big[ \mathbbm{1}\big\{ \widehat{W}_{\a} \ne W_\a\big\} \big\vert \Ac, K_\a = k_\a\big]$}
			We denote by $\{k_\a \to k_\a'\}$ the event that $K_\a = k_\a$ and the initial estimation of $K_\a$ in~\eqref{eq:Ka_estimation} outputs $k_\a'$. According to~\eqref{eq:Ka_estimation}, 
			\begin{align}
				&\P{k_\a \to {k}'_\a} \notag \\ &= \PP\big[m_\a(\vec Y_{\a},{k}_{\a}') > m_\a(\vec Y_{\a},k), \forall k \in [k_{\a,\ell}:k_{\a,u}] \setminus \{{k}_{\a}'\} \notag \\
                &\qquad \qquad \;\vert\; K_\a = k_\a\big] \\
				&\le \min_{k \in \{0\} \cup [k_{\a,\ell}:k_{\a,u}] \setminus \{{k}_{\a}'\}} \P{m_\a(\vec Y_{\a},{k}_{\a}') > m_\a(\vec Y_{\a},k)}. \label{eq:tmp1804}
			\end{align}
			Under the new measure, $\vec Y_{\a} \sim \mathcal{N}(\mathbf{0}, (1+k_{\a}\sf{P}_\a')\vec I_{\sf{n}_{\a}})$. Thus the right-hand side of~\eqref{eq:tmp1804} is given by $\zeta(k_\a, {k}_\a')$ defined in~\eqref{eq:zeta}.
			
			Given $k_\a > 0$ and $k_\a \to k_\a'$, an AMD $\{\widehat{W}_\a \ne W_\a\}$ occurs if $k_\a' = 0$, or if $k_\a' \in [k_{\a,\ell}:k_{\a,u}]$ but the decoder~\eqref{eq:alarm_decoding} returns the wrong alarm message.  
			The latter event occurs if  some scaled version of a wrong alarm codeword is closer to the received signal than the correct alarm codeword,~i.e.,  
			\begin{equation}
				\|\vec y_\a -  \hat{k}_{\a} \revise{\vec C_{\a, \widehat{w}}} \|^2 <\norm{\vec y_\a - {k}_\a \revise{\vec C_{\a, W_\a}}},
				\label{eq:slicingAlarmErrorEvent1}
			\end{equation} 
			for some $\hat{k}_\a \in \{0\} \cup [k_{\a,\ell} : k_{\a,u}]$ and $\widehat{w} \ne W_\a$. 
			We denote by $F_\amd(\hat{k}_{\a})$ the set of $\widehat w$ such that~\eqref{eq:slicingAlarmErrorEvent1} holds for a given $\hat{k}_\a$. It follows that $\E[\Cc_\a]{ \ind{\widehat{W}_{\a} \ne W_\a} \big\vert \Ac, K_\a = k_\a}$ is bounded as
			\begin{align}
				&\E[\Cc_\a]{ \ind{\widehat{W}_{\a} \ne W_\a} \big\vert \Ac, K_\a = k_\a}  \notag \\
                &\le \P{k_\a \to 0} 
                \notag \\
				&\quad 
				+ \P{k_\a \!\to\! k_\a', k_\a' \in [k_{\a,\ell}:k_{\a,u}]}  \!\!\sum_{\hat{k}_\a \in \{0\} \cup [k_{\a,\ell} : k_{\a,u}]} \!\!\!\!\!\P{F_\amd(\hat{k}_{\a})} \\
				&\le\zeta(k_\a, 0) \notag \\
                &\quad +\min\Bigg\{1,\sum_{k_\a' =k_{\a,\ell}}^{k_{\a,u}} \zeta(k_\a, k_\a')  \Bigg\} \!\sum_{\hat{k}_\a \in  \{0\} \cup [k_{\a,\ell} : k_{\a,u}]} \!\!\!\P{F_\amd(\hat{k}_{\a})}\!. \label{eq:tmp74148}
			\end{align}
			
			\subsubsection{Bounding $\P{F_\amd(\hat{k}_{\a})}$ Via the \gls{RCUs}}
			It remains to bound the probability $\P{F_\amd(\hat{k}_{\a})} = \P{\bigcup_{\widehat{w} \in \Mc_\a \setminus \{W_\a\}} \big\{\big\|\vec Y_\a -  \hat{k}_{\a}\revise{\vec C_{\a, \widehat{w}}} \big\|^2 <\norm{\vec Y_\a - {k}_\a\revise{\vec C_{\a, W_\a}}}\big\}}.$
			We do this using the \gls{RCUs}~\cite[Th. 16]{Polyanskiy2010},~\cite{Martinez2011RCUs}.
			Specifically, by applying a tightened version of the union bound, we obtain
			\begin{align}
				&\P{F_\amd(\hat{k}_{\a})} \notag \\
                &\le \EE\Big[\min\Big\{ 1, (\sf{M}_\a - 1) \notag \\
                &\quad \cdot \P{\big\| \vec Y' - \hat{k}_{\a}\widehat{\vec X}\big\|^2<\norm{\vec Y' - {k}_{\a}{\vec X'}} \big\vert \vec X', \vec Y'} \Big\}\Big] \label{eq:tmp74}
			\end{align}
			where $\{\vec Y',\vec X', \widehat{\vec X}\}$ has the same joint distribution as $\{\vec Y_\a,\revise{\vec C_{\a,W_\a}},\revise{\vec C_{\a,\widehat{w}}}\}$. That is, $\vec{X}' = [X'_1 \dots X'_{\sf{n}_{\a}}]^\T$ follows $\mathcal{N}(\mathbf{0},\sf{P}'_\a\vec I_{\sf{n}_{\a}})$; given $X'_i = x_i'$, we have that $\vec Y' = [Y'_1 \dots Y'_{\sf{n}_{\a}}]^\T$ with $Y'_i \sim \mathcal{N}({k}_{\a}x'_i,1)$; and $\widehat{\vec X} = [\widehat X'_1 \dots \widehat X'_{\sf{n}_{\a}}]^\T\sim \mathcal{N}(\mathbf{0},\sf{P}'_\a \vec I_{\sf{n}_{\a}})$, independent of both $\vec X'$ and $\vec Y'$. Next, by applying the Chernoff bound in Lemma~\ref{lem:Chernoff}, we obtain that
			\begin{align}
				&\P{\big\|\vec Y' - \hat{k}_{\a}\widehat{\vec X}\big\|^2 < \norm{ \vec Y' - k_{\a}\vec X'} \big\vert \vec X', \vec Y'} \notag \\
                &\quad \le \frac{\EE_{\widehat{\vec X}}\big[\exp\big(-s \| \vec Y' - \hat{k}_\a \widehat{\vec X}\|^2\big)\big]}{\exp\left(-s \| \vec Y' - {k}_\a \vec X'\|^2\right)} \label{eq:tmp79}
			\end{align} 
			for every  $s > 0$. Substituting~\eqref{eq:tmp79} into~\eqref{eq:tmp74}, we obtain that, for every  $s > 0$,
			\begin{align}
				&\P{F_\amd(\hat{k}_{\a})} \notag \\
                &\le \EE\Bigg[ \min \Bigg\{1, \exp\Bigg(\ln (\sf{M}_\a - 1) \notag \\
                &\qquad + \ln \frac{\EE_{\widehat{\vec X}}\big[\exp\big(-s \| \vec Y' - \hat{k}_\a \widehat{\vec X}\|^2\big)\big]}{\exp\left(-s \| \vec Y' - {k}_\a \vec X'\|^2\right)} \Bigg)\Bigg\}\Bigg] \\
				&= \EE\Bigg[ \min \Bigg\{1, \exp\Bigg(\ln (\sf{M}_\a - 1) \notag \\
                &\qquad- \sum_{i=1}^{\sf{n}_\a} \ln \frac{\exp\left(-s (Y'_i \!-\! {k}_\a X'_i)^2\right)}{\EE_{\widehat{X}_i}\big[\exp\big(\!-\!s (Y'_i \!-\! \hat{k}_\a \widehat{X}_i)^2\big)\big]}\Bigg)\Bigg\}\Bigg]. \label{eq:tmp85}
			\end{align}
			We define the generalized information density as
			$
				\imath_s(\hat{k}_\a, X'_i; Y'_i) \triangleq \ln \frac{\exp\left(-s ( Y'_i - {k}_\a X'_i)^2\right)}{\E[\widehat{X}_i]{\exp(-s (Y'_i - \hat{k}_\a \widehat{X}_i)^2)}}.
			$
			After some manipulations using~\eqref{eq:chi2_MGF} in Lemma~\ref{lem:chi2}, we deduce that $\imath_s(\hat{k}_\a, x; y)$ can be written as in~\eqref{eq:def_gen_info_den}. Using~$\imath_s(\hat{k}_\a, X'_i; Y'_i)$, we can rewrite~\eqref{eq:tmp85}, upon optimizing over $s$, as 
			\begin{align}
				\P{F_\amd(\hat{k}_{\a})} &\le \min_{s>0} \EE\Bigg[\min \Bigg\{1, \exp\bigg(\ln (\sf{M}_\a - 1) \notag \\
                &\qquad \qquad -\sum_{i=1}^{\sf{n}_\a} \imath_s(\hat{k}_\a, X'_i; Y'_i) \bigg)\Bigg\}\Bigg]. \label{eq:tmp99154}
			\end{align}
			Next, by observing that, for every positive random variable $Q$, it holds that $\E{\min\{1,Q\}} = \P{Q \ge V}$ where $V$ is uniformly distributed on $[0,1]$, we obtain that 
			\begin{align}
				\P{F_\amd(\hat{k}_{\a})} \le \min_{s>0} \P{\sum_{i=1}^{\sf{n}_\a} \imath_s(\hat{k}_\a, X'_i; Y'_i) \le \ln\frac{\sf{M}_\a-1}{V} }. \label{eq:tmp103144}
			\end{align}
			Finally, by substituting~\eqref{eq:tmp103144} into~\eqref{eq:tmp74148} and~\eqref{eq:tmp74148} into~\eqref{eq:tmp24}, we complete the proof.

			\subsection{Proof of the Bound~\eqref{eq:RCU_AFP} on $P_\afp$} 
			\label{proof:OMA_FP}
			Consider the case where no alarm event occurs (thus $k_\a = 0$) and the estimation step outputs $k_\a'$. An AFP occurs if $K_\a' > 0$ in the initial estimation step~\eqref{eq:Ka_estimation} and $\widehat{K}_\a > 0$ in the message decoding step~\eqref{eq:alarm_decoding}. The latter event occurs if an alarm codeword, scaled by some $\hat{k}_\a \in[k_{\a,\ell} : k_{\a,u}]$, is closer to the received signal, which consists just of additive noise, than the all-zero codeword, i.e.,  
			\begin{equation}
				\|\hat{k}_\a\revise{\vec C_{\a, \widehat{w}}}-\vec Z_{\mathrm{a}}\|^2 < \norm{\vec Z_{\mathrm{a}}}
				\label{eq:tmp129}
			\end{equation}
			for some $\widehat{w}\in \Mc_\a$.
			Let $F_{\afp}(\hat{k}_{\a})$ denote the set of $\widehat{w}$ such that~\eqref{eq:tmp129} holds for a given $\hat{k}_\a$. 
			%
			Then, the AFP probability is given by
			\begin{align}
				P_{\afp} &= \sum_{k_\a' = k_{\a,\ell}}^{k_{\a,u}}  \!\P{0 \to {k}_\a'} \sum_{\hat{k}_\a = k_{\a,\ell}}^{k_{\a,u}} \P{F_{\afp}(\hat{k}_{\a})} \\
				&\le \min\Bigg\{1,\sum_{k_\a' = k_{\a,\ell}}^{k_{\a,u}} \zeta(0, {k}_\a') \Bigg\} \sum_{\hat{k}_\a =k_{\a,\ell}}^{k_{\a,u}} \P{F_{\afp}(\hat{k}_{\a})}. \label{eq:tmp194}
			\end{align}
			Next, we use again the \gls{RCUs} to bound $\P{F_{\afp}(\hat{k}_{\a})}$. Specifically, we first apply a tightened version of the union bound to obtain
			\begin{align}
				&\P{F_{\afp}(\hat{k}_{\a})} \notag \\
                &\le \E[\vec Z_\a]{\min\left\{1, \sf{M}_\a \P{\big\|\vec Z_\a - \hat{k}_\a \widehat{\vec X}\big\|^2 < \norm{{\vec Z}_\a}} \right\}} \label{eq:tmp201}
			\end{align}
			where $\widehat{\vec X}$ is identically distributed to $\revise{\vec C_{\a,\widehat{w}}}$, i.e., $\widehat{\vec X} \sim \Nc(\mathbf{0},\sf{P}'_\a\vec I_{\sf{n}_\a})$. We then apply the Chernoff bound in Lemma~\ref{lem:Chernoff}, and conclude that, for every $s>0$,
			\begin{align}
				&\P{\big\|\vec Z_\a - \hat{k}_\a \widehat{\vec X}\big\|^2 < \norm{\vec Z_\a}} \notag \\
				&\le e^{s \norm{\vec Z_\a}} \E[\widehat{\vec X}]{\exp\left(-s\big\|\vec Z_\a - \hat{k}_\a \widehat{\vec X}\big\|^2\right)} \label{eq:tmp1916}  \\
				&= e^{s \norm{\vec Z_{\a}}_2}\frac{\exp\Big(-\frac{s\norm{\vec Z_{\a}}_2}{1+2\hat{k}_{\a}^2\sf{P}'_\a s }\Big)}{(1+2\hat{k}_{\a}^2\sf{P}'_\a s )^{\sf{n}_{\a}/2}}\label{eq:h-omaFalseFirstIdentity}\\
				&= \exp\Big(\beta\norm{\vec Z_{\a}}_2-\frac{\sf{n}_{\a}}{2}\ln\big(1+2\hat{k}_{\a}^2\sf{P}'_\a s \big)\Big)
				\label{eq:tmp205}
			\end{align} 
			with $\beta \triangleq s(1 - (1+2\hat{k}_{\a}^2\sf{P}'_\a s)^{-1})$. Here, to obtain~\eqref{eq:h-omaFalseFirstIdentity}, we computed the expectation in~\eqref{eq:tmp1916} using the identity~\eqref{eq:chi2_MGF} in Lemma~\ref{lem:chi2}. 
			Substituting~\eqref{eq:tmp205} into~\eqref{eq:tmp201}, we obtain
			\begin{align}
				&\P{F_{\afp}(\hat{k}_{\a})} \notag \\
                &\leq \E[\vec Z_{\a}]{\min\left\{1,\sf{M}_\a \exp\Big(\beta\norm{\vec Z_{\a}}\!-\!\frac{\sf{n}_{\a}}{2}\ln\big(1\!+\!2\hat{k}_{\a}^2\sf{P}'_\a s \big)\Big)\right\}} \label{eq:tmp252176}\\
				&= \P{\sf{M}_\a \exp\Big(\beta\norm{\vec Z_{\a}}-\frac{\sf{n}_{\a}}{2}\ln\big(1+2\hat{k}_{\a}^2\sf{P}'_\a s \big)\Big) \le V} \\
				&= \P{\norm{\vec Z_{\a}} \ge \beta^{-1} \left(\frac{\sf{n}_{\a}}{2}\ln(1+2\hat{k}_{\a}^2\sf{P}'_\a s ) - \ln \frac{\sf{M}_\a}{V}\right)} \label{eq:tmp2353}
			\end{align}
			where $V$ is uniformly distributed on $[0,1]$.
			Recall that $\vec Z_\a \sim \Nc(\mathbf{0}, \vec I_{\sf{n}_\a})$. By applying~\eqref{eq:chi2_CCDF} in Lemma~\ref{lem:chi2} to~\eqref{eq:tmp2353}, we obtain that
			\begin{align}
				&\P{F_{\afp}(\hat{k}_{\a})} \leq \min_{s>0} \EE_V\Bigg[\frac{1}{\Gamma(\sf{n}_\a/2)} \notag \\
                &\qquad \cdot \Gamma\left(\frac{\sf{n}_\a}{2}, \frac{1}{2\beta} \bigg(\frac{\sf{n}_{\a}}{2}\ln(1+2\hat{k}_{\a}^2\sf{P}'_\a s ) - \ln \frac{\sf{M}_\a}{V}\bigg)\right)\Bigg]. \label{eq:tmp1932}
			\end{align} 
			Finally, we substitute~\eqref{eq:tmp1932} into~\eqref{eq:tmp194} to complete the proof.
			
			
			\section{Proof of Theorem~\ref{th:HNOMA}}\label{proof:HNOMA}
			
			\subsection{Proof of the Bounds~\eqref{eq:bar_eps_AMD} and~\eqref{eq:bar_eps_AFP} on $P_\amd$ and $P_\afp$}
			\label{proof:NOMA_alarm}
			The AMD probability is computed as 
				$\P{\widehat{W}_{\a} \ne W_\a \vert \Ac} = \E[K_\a, {K}_\s, \Cc_\a, \Cc_\s]{\ind{\widehat{W}_{\a} \ne W_\a} \big\vert \Ac}$,
			where $K_\a \sim {\rm Bino}(\sf{K},\rho_\d)$ is the number of alarm users and $K_\s \sim {\rm Bino}(\sf{K},\rho_\s)$ is the number of standard users. 
			We change the measure to the one for which i) $\vec X_\a = \vec C_{\a,W_\a}$ instead of $\vec X_\a = \vec C_{\a,W_\a} \ind{\|\vec C_{\a, W_\a}\|^2 \leq \sf{n} \sf{P}_\a}$, ii) $K_\a \in [k_{\a,\ell}:k_{\a,u}]$, iii) $\vec X_{\s,k} = \vec C_{\s,W_{\s,k}}$ instead of $\vec X_{\s,k} = \vec C_{\s,W_{\s,k}} \ind{\|\vec C_{\s, W_{\s,k}}\|^2 \leq \sf{n} \sf{P}_\s}$, iv) ${K}_\s \in [{k}_{\s,\ell}:{k}_{\s,u}]$, and v) the transmitted standard codewords are distinct. This is done at a cost of adding a constant bounded by $\nu_2$, given in~\eqref{eq:p2}, to the original expectation. 
			Under the new measure, the only difference with Appendix~\ref{proof:OMA_AMD} is that $\sf{n}_\a$ is replaced by $\sf{n}$, and the equivalent noise is now $\overline{\vec Z} = \sum_{k = 1}^{K_\s} \vec X_{\s,k} + \vec Z \sim \Nc(\mathbf{0}, (1+{K}_\s \sf{P}'_\s)\vec I_{\sf{n}})$.  Given $K_\s = k_\s$, by adapting the steps in Appendix~\ref{proof:OMA_AMD} to codeword length $\sf{n}$ and noise variance $(1+k_\s P')$, we bound $\E[K_\a, \Cc_\a, \Cc_\s]{\ind{\widehat{W}_{\a} \ne W_\a} \big\vert \Ac, {K}_\s = k_\s}$ by $\bar{\epsilon}_\amd$ given in~\eqref{eq:bar_eps_AMD}. Finally, by averaging over $K_\s$, we conclude that $\P{\widehat{W}_{\a} \ne W_\a \big\vert \Ac} \le \sum_{{k}_\s = {k}_{\s,\ell}}^{{k}_{\s,u}} P_{{K}_\s}( k_\s) \bar{\epsilon}_\amd(k_\s) + \nu_2$.
			
			
			The AFP probability is computed as $\P{\widehat{W}_{\a} \ne w_\e \big \vert \bar{\Ac}} = \E[{K}_\s, \Cc_\a, \Cc_\s]{\ind{\widehat{W}_{\a} \ne w_\e} \big \vert \bar{\Ac}}$. We change the measure to the one for which i) $\vec X_{\s,k} = \vec C_{\s,W_{\s,k}}$ instead of $\vec X_{\s,k} = \vec C_{\s,W_{\s,k}} \mathbbm{1}\{\|\vec C_{\s, W_{\s,k}}\|^2 \leq \sf{n} \sf{P}_\s\}$, ii) ${K}_\s \in [k_{\s,\ell}:k_{\s,u}]$, and iii) the transmitted standard messages are distinct. This is done at a cost of adding a constant bounded by $\nu_3$ given in~\eqref{eq:p3}. From here, the only difference with Appendix~\ref{proof:OMA_FP} is that, in the case of no alarm, the received signal is $\vec Y = \sum_{k=1}^{{K}_\s} \vec X_{\s,k} + \vec Z \sim \Nc(\mathbf{0}, (1+ K_\s \sf{P}'_\s)\vec I_{\sf{n}})$ instead of just noise. Given ${K}_\s = k_\s$, by adapting the steps in Appendix~\ref{proof:OMA_FP} to this output distribution, we bound $\E[\Cc_\a, \Cc_\s]{\ind{\widehat{W}_{\a} \ne w_\e} \big \vert \bar{\Ac}, {K}_\s = k_\s}$ by $\bar{\epsilon}_\afp$ given in~\eqref{eq:bar_eps_AFP}. Finally, by averaging over $K_\s$, we conclude that $\P{\widehat{W}_{\a} \ne w_\e \big \vert \bar{\Ac}}  \le \sum_{{k}_\s = {k}_{\s,\ell}}^{{k}_{\s,u}} P_{{K}_\s}( k_\s) \bar{\epsilon}_\afp(k_\s) + \nu_3$.
			\subsection{Proof of the Bounds~\eqref{eq:RCU_NOMA_SMD_noAlarm} and \eqref{eq:RCU_NOMA_SFP_noAlarm} on $P_{\smd \vert \bar{\Ac}}$ and $P_{\sfp \vert \bar{\Ac}}$} \label{proof:NOMA_std_noAlarm}
			
		For convenience, we set $E_\smd \triangleq \frac{1}{|\widetilde{\Wc}_\s|}\sum_{i=1}^{|\widetilde{\Wc}_\s|}\P{ \widetilde{W}_{\s,i} \notin \widehat{\Wc}_\s}$. 
		We use the law of total probability to expand and then bound $P_{\smd \vert \bar{\Ac}}$ as follows:
		\begin{align}
			&P_{\smd \vert\bar{\Ac}} \notag \\ 
            &= \E{E_\smd \vert \bar{\Ac}} \label{eq:tmp2018}\\
			&= \E{E_\smd \vert \bar{\Ac}, \widehat{W}_{\a} \ne w_\e } \P{\widehat{W}_{\a} \ne w_\e \vert \bar{\Ac}}  \notag \\
            &\quad + \E{E_\smd \vert \bar{\Ac},  \widehat{W}_{\a} = w_\e} \P{ \widehat{W}_{\a} = w_\e \vert \bar{\Ac}} \\
			&\le \P{\widehat{W}_{\a} \ne w_\e \vert \bar{\Ac}} \notag \\
            &\quad + \Big(1-\P{\widehat{W}_{\a} \ne w_\e \vert \bar{\Ac}}\Big) \E{E_\smd \vert \bar{\Ac}, \widehat{W}_{\a} = w_\e  }  \label{eq:tmp65104}\\
			&= 1 - \Big(1- \P{\widehat{W}_{\a} \ne w_\e \vert \bar{\Ac}}\!\Big)\Big(1- \E{E_\smd \vert \bar{\Ac}, \widehat{W}_{\a} = w_\e} \!\Big) \label{eq:tmp2021} \\
			&\le 1 - \!\sum_{k_\s = k_{\s,\ell}}^{k_{\s,u}} \!P_{K_\s}(k_\s) \Big(1- \P{\widehat{W}_{\a} \ne w_\e \vert \bar{\Ac},K_\s = k_\s}\Big)\notag \\
            &\qquad\qquad  \cdot \Big(1- \E{E_\smd \vert \bar{\Ac}, \widehat{W}_{\a} = w_\e, K_\s = k_\s} \Big) \notag \\
			&\quad + \nu_2. \label{eq:tmp2083}
		\end{align}
		Here, we obtained~\eqref{eq:tmp65104} by assuming that one cannot decode the standard messages if an AFP occurs. In~\eqref{eq:tmp2083}, we made the same change of measure as in the bound of $P_\amd$, and the probability and expectation therein are computed with respect to the new measure. 
		In Appendix~\ref{proof:NOMA_alarm}, we have bounded $\P{\widehat{W}_{\a} \ne w_\e \vert \bar{\Ac},K_\s \!=\! k_\s} = \E[\Cc_\a, \Cc_\s]{\ind{\widehat{W}_{\a} \ne w_\e} \big \vert \bar{\Ac}, {K}_\s = k_\s}$ by $\bar{\epsilon}_\afp(k_\s)$.  
		Furthermore, when there is no alarm and no AFP, the standard message list is decoded as in the standard block of \gls{HOMA}. 
		Therefore, $\E{E_\smd \vert \bar{\Ac}, \widehat{W}_{\a} = w_\e, K_\s = k_\s}$ 
		is upper-bounded by $\bar{\epsilon}_\smd$ 
		in Theorem~\ref{th:RCU_HOMA}, adapted to the codeword length $\sf{n}$. 
		Finally, using these bounds in~\eqref{eq:tmp2083}, we conclude that $P_{\smd \vert \bar{\Ac}} \le \epsilon_{\smd \vert \bar{\Ac}}$ with $\epsilon_{\smd \vert \bar{\Ac}}$ given in~\eqref{eq:RCU_NOMA_SMD_noAlarm}. The bound $\epsilon_{\sfp \vert \bar{\Ac}}$ of $P_{\sfp \vert \bar{\Ac}}$, given in~\eqref{eq:RCU_NOMA_SFP_noAlarm}, follows similarly.

		\subsection{Proof of the Bounds~\eqref{eq:RCU_NOMA_SMD_alarm} and~\eqref{eq:RCU_NOMA_SFP_alarm} on $P_{\smd \vert \Ac}$ and $P_{\sfp \vert \Ac}$} \label{proof:NOMA_std_alarm}
		
		By following similar steps as in~\eqref{eq:tmp2018}--\eqref{eq:tmp2083}, we obtain that
		\begin{align}
			P_{\smd \vert \Ac} &\le 1 - \sum_{k_\s = k_{\s,\ell}}^{k_{\s,u}} \!P_{K_\s}(k_\s) \Big(1\!-\! \P{\widehat{W}_{\a} \ne W_\a \vert \Ac,K_\s = k_\s}\!\Big)\notag \\
            &\qquad \qquad \cdot \Big(1- \E{E_\smd \vert \Ac, \widehat{W}_{\a} \!=\! W_\a, K_\s = k_\s} \!\Big) \notag \\
			&\quad + \nu_2. \label{eq:tmp2114}
		\end{align}
		In Appendix~\ref{proof:NOMA_alarm}, we have bounded $\mathbb{P}\Big[\widehat{W}_{\a} \!\ne\! W_\a \vert \Ac,K_\s \!=\! k_\s\Big] = \mathbb{E}_{K_\a, \Cc_\a, \Cc_\s}\Big[\mathbbm{1}\{\widehat{W}_{\a} \!\ne\! W_\a\} \big\vert \Ac, {K}_\s = k_\s\Big]$ by $\bar{\epsilon}_\amd(k_\s)$. We next bound the term  $\E{E_\smd \vert \Ac, \widehat{W}_{\a} = W_\a, K_\s = k_\s}$ 
		as
		\begin{align}
			&\E{E_\smd \vert \Ac, \widehat{W}_{\a} = W_\a, K_\s = k_\s} \notag \\
			&\le \sum_{k_\a = k_{\a,\ell} }^{k_{\a,u}} P_{K_\a}(k_\a)
			 \sum_{\hat{k}_\a \in \{0\} \cup [k_{\a,\ell}:k_{\a,u}]} \notag \\
            &\quad  \P{\widehat{K}_\a = \hat{k}_\a \vert \widehat{W}_{\a} = W_\a, K_\a = k_\a, K_\s = k_\s} \notag \\
			&\qquad \cdot \E{E_\smd \vert \Ac, \widehat{W}_{\a} = W_\a, \widehat{K}_\a = \hat{k}_\a, K_\a = k_\a, K_\s = k_\s}. \label{eq:tmp2442} 
		\end{align}
		Given $(\widehat{W}_{\a},\widehat{K}_\a, K_\a) = (W_\a, \hat{k}_\a, k_\a)$, the residual interference plus noise after interference cancellation is $(k_\a - \hat{k}_\a) \vec C_{\a,W_\a} + \vec Z \sim \Nc(\mathbf{0}, (1 + (k_\a - \hat{k}_\a)^2 \sf{P}_\a')\vec I_{\sf{n}})$. The receiver decodes the standard messages under this effective noise. Therefore, $\E{E_\smd \vert \Ac, \widehat{W}_{\a} = W_\a, \widehat{K}_\a = \hat{k}_\a, K_\a = k_\a, K_\s = k_\s}$ is bounded by~$\bar{\epsilon}_\smd$ in Theorem~\ref{th:RCU_HOMA}, adapted to codeword length $\sf{n}$ and noise variance $1 + (k_\a - \hat{k}_\a)^2 \sf{P}_\a'$. 
		
		It remains to bound $\P{\widehat{K}_\a \!=\! \hat{k}_\a \vert \widehat{W}_{\a} \!=\! W_\a, K_\a \!=\! k_\a, K_\s \!=\! k_\s}$. Given 
		$(\widehat{W}_{\a},{K}_\a, K_\s) \!=\! (W_\a, {k}_\a, k_\s)$, if $\widehat{K}_\a = \hat{k}_\a > 0$,  it must hold that i) the estimation step returned $k'_\a \in [k_{\a,\ell}:k_{\a,u}]$, and ii) $\|\vec Y - \hat{k}_\a \vec C_{\a, W_\a}\|^2 \le \|\vec Y - {k}_\a \vec C_{\a, W_\a}\|^2$ or, equivalently,
		$\|(k_\a -  \hat{k}_\a) \vec C_{\a, W_\a} + \overline{\vec Z} \|^2 \le  \|\overline{\vec Z} \|^2$, where $\overline{\vec Z} = \sum_{k = 1}^{{k}_\s} \vec X_{\s,k} + \vec Z \sim \Nc(\mathbf{0}, (1+{k}_\s \sf{P}'_\s)\vec I_{\sf{n}})$. If $\widehat{K}_\a = \hat{k}_\a = 0$, either both conditions above hold, or the estimation step returned $k_\a' = 0$.
		Therefore, 
		\begin{align}
			&\P{\widehat{K}_\a = \hat{k}_\a \vert \widehat{W}_{\a} = W_\a, K_\a = k_\a, K_\s = k_\s} \notag \\
			&\le \mathbbm{1}\{\hat{k}_\a = 0\}\P{k_\a \to 0} \notag \\
            &\quad + \P{k_\a \to k'_\a, k_\a' \in [k_{\a,\ell}:k_{\a,u}]} \notag \\
            &\qquad \cdot \P{\|(k_\a -  \hat{k}_\a) \vec C_{\a, W_\a} + \overline{\vec Z} \|^2 \le  \|\overline{\vec Z} \|^2}  \\
			&\le  \mathbbm{1}\{\hat{k}_\a = 0\}\eta(k_\a,0,k_\s) \notag \\
            &\quad + \min\Bigg\{1,\sum_{k_\a' = k_{\a,\ell}}^{k_{\a,u}}  \eta(k_\a, k'_\a, {k}_\s) \Bigg\} \notag \\
            &\qquad \cdot\P{\|(k_\a -  \hat{k}_\a) \vec C_{\a, W_\a} + \overline{\vec Z} \|^2 \le  \|\overline{\vec Z} \|^2}.  \label{eq:tmp97356}
		\end{align}
		Using the Chernoff bound, we bound $\P{\|(k_\a \!-\!  \hat{k}_\a) \vec C_{\a, W_\a} + \overline{\vec Z} \|^2 \le  \|\overline{\vec Z} \|^2}$ by $\Big(1+\frac{(k_\a - \hat{k}_\a)^2 \sf{P}'_\a}{4(1+{k}_\s \sf{P}'_\s)}\Big)^{\!-\sf{n}/2}$. 
		Substituting this 
        into~\eqref{eq:tmp97356}, then \eqref{eq:tmp97356} into~\eqref{eq:tmp2442}, and finally~\eqref{eq:tmp2442} into~\eqref{eq:tmp2114}, we conclude that $P_{\smd \vert \Ac} \le \epsilon_{\smd \vert \Ac}$, with $\epsilon_{\smd \vert \Ac}$ given in~\eqref{eq:RCU_NOMA_SMD_alarm}. The proof that $P_{\sfp \vert \Ac} \le \epsilon_{\sfp \vert \Ac}$, with $\epsilon_{\sfp \vert \Ac}$ given in~\eqref{eq:RCU_NOMA_SFP_alarm}, follows similarly.
		



		\bibliographystyle{IEEEtran}
		\bibliography{mybib.bib}
		
		%
		%
		%
		  
		
		

        \begin{IEEEbiography}[{\includegraphics[width=1in,height=1.25in,clip,keepaspectratio]{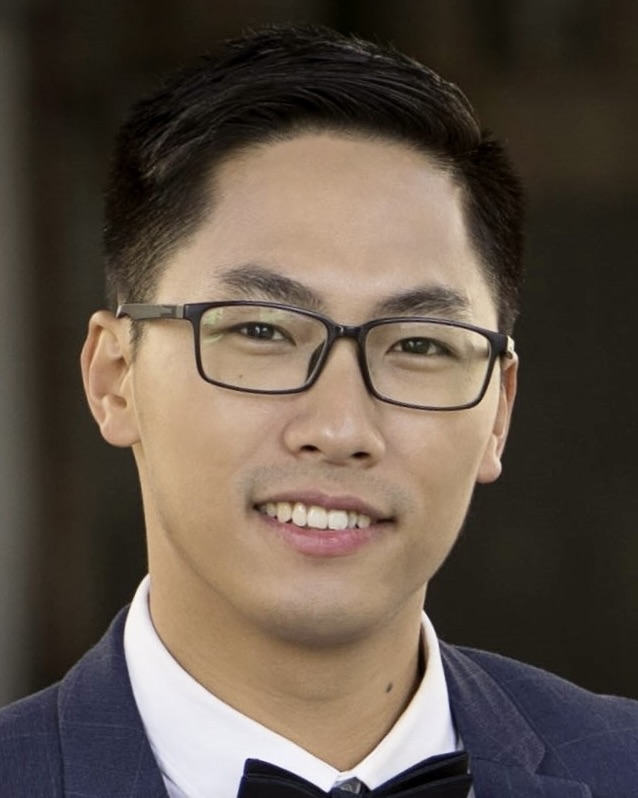}}]{Khac-Hoang Ngo}
	(Member, IEEE) received the B.E. degree (Hons.) in electronics and telecommunications from University of Engineering and Technology, Vietnam National University, Hanoi, Vietnam, in 2014; and the M.Sc. degree (Hons.) and Ph.D. degree in wireless communications from CentraleSupélec, Paris-Saclay University, France, in 2016 and 2020, respectively. His Ph.D. thesis was also realized at Paris Research Center, Huawei Technologies France. Since September 2020, he has been a postdoctoral researcher at Chalmers University of Technology, Sweden. He is also an adjunct lecturer at University of Engineering and Technology, Vietnam National University Hanoi, Vietnam. His research interests include wireless communications, information theory, and decentralized learning, with an emphasis on massive random access, privacy of federated learning, age of information, MIMO, and noncoherent communications. He received the ``Signal, Image \& Vision Ph.D. Thesis Prize'' by Club EEA, GRETSI and GdR-ISIS, France, and the Marie Skłodowska-Curie Actions (MSCA) Individual Fellowship in 2021.
\end{IEEEbiography}

\begin{IEEEbiography}[{\includegraphics[width=1in,height=1.25in,clip,keepaspectratio]{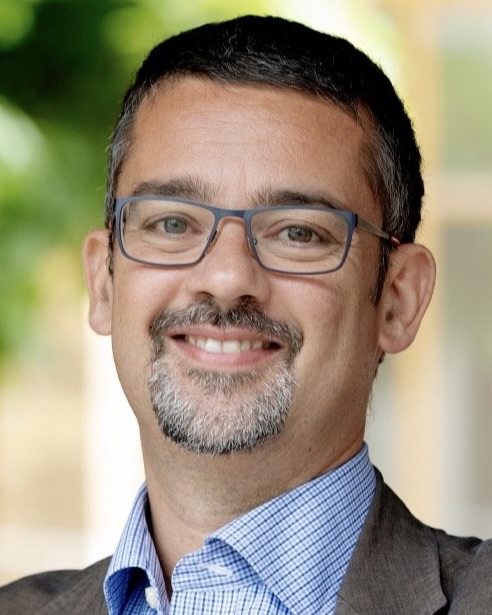}}]{Giuseppe Durisi}
    (Senior Member, IEEE) received the Laurea (summa cum laude) and Ph.D. degrees from the Politecnico di Torino, Italy, in 2001 and 2006, respectively.
    From 2002 to 2006, he was with the Istituto Superiore Mario Boella, Turin, Italy. From 2006 to 2010, he was a Post-Doctoral Researcher at ETH Zurich, Zürich, Switzerland. In 2010, he joined the Chalmers University of Technology, Gothenburg, Sweden, where he is currently a Professor with
    the Communication Systems Group. His research interests are in the areas of communication and information theory and machine learning. He is the recipient of the 2013 IEEE ComSoc Best Young Researcher Award for the Europe, Middle East, and Africa region, and is coauthor of a paper that won the Student Paper Award at the 2012 International Symposium on Information Theory, and of a paper that won the 2013 IEEE Sweden VT-COM-IT Joint Chapter Best Student Conference Paper Award. From 2011 to 2014, he served as a publications editor for the IEEE TRANSACTIONS ON INFORMATION THEORY. From 2015 to 2021, he served as associate editor for the  IEEE TRANSACTIONS ON COMMUNICATIONS.
\end{IEEEbiography}

\begin{IEEEbiography}[{\includegraphics[width=1in,height=1.25in,clip,keepaspectratio]{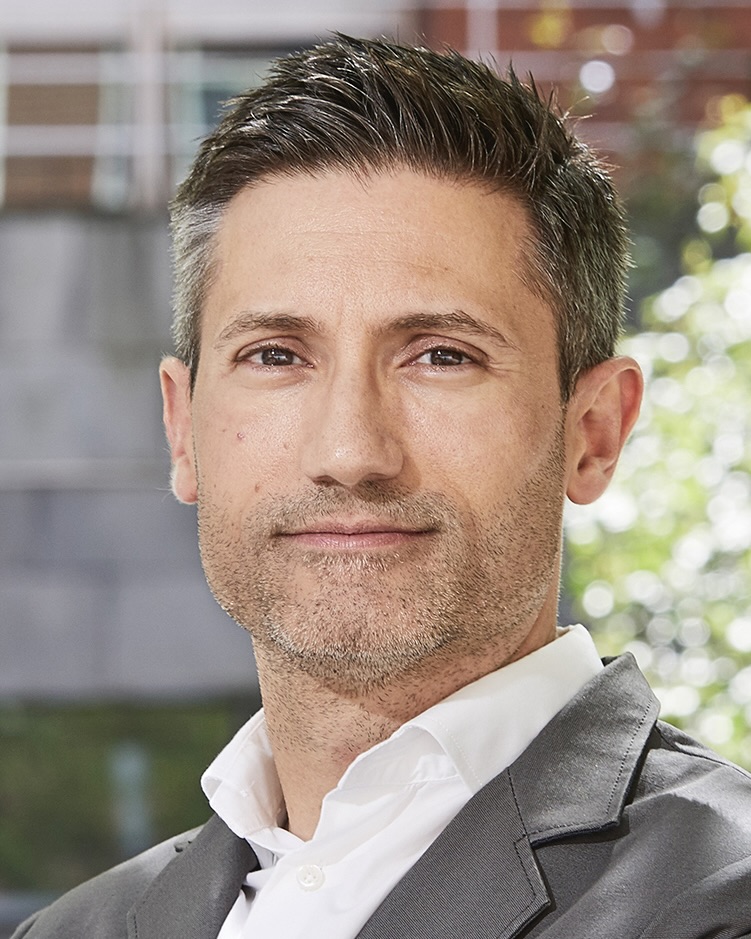}}]{Alexandre Graell i Amat}
    (Senior Member, IEEE) received the M.Sc. and Ph.D. degrees in electrical engineering from the Politecnico di Torino, Turin,
    Italy, in 2000 and 2004, respectively, and the M.Sc. degree in telecommunications engineering from the Universitat Politècnica de Catalunya, Barcelona, Catalonia, Spain, in 2001. From 2001 to 2002, he was a Visiting Scholar with the University of California at San Diego, La Jolla, CA,
    USA. From 2002 to 2003, he held a visiting appointment at Universitat Pompeu Fabra, Barcelona, and the Telecommunications Technological Center of Catalonia, Barcelona. From 2001 to 2004, he held a part-time appointment
    at STMicroelectronics Data Storage Division, Milan, Italy, as a Consultant on coding for magnetic recording channels. From 2004 to 2005, he was a Visiting Professor with Universitat Pompeu Fabra. From 2006 to 2010, he was with the Department of Electronics, IMT Atlantique (formerly ENST Bretagne), Brest, France. Since 2019, he has also been an Adjunct Research Scientist with Simula UiB, Bergen, Norway. He is currently a Professor with the Department of Electrical Engineering, Chalmers University of Technology,     Gothenburg, Sweden. His research interests are in the field of coding theory with application to distributed learning and computing, storage, privacy and security, and communications. He received the Marie Skłodowska-Curie Fellowship from the European Commission and the Juan de la Cierva Fellowship from the Spanish Ministry of Education and Science. He received the IEEE Communications Society 2010 Europe, Middle East, and Africa Region Outstanding Young Researcher Award. He was the General Co-Chair of the 7th International Symposium on Turbo Codes and Iterative Information Processing, Sweden, in 2012, and the TPC Co-Chair of the 11th International Symposium on Topics in Coding, Canada, in 2021. He was an Associate Editor of the IEEE COMMUNICATIONS LETTERS from 2011 to 2013. He was an Associate Editor and the Editor-at-Large of the IEEE TRANSACTIONS     ON COMMUNICATIONS from 2011 to 2016 and 2017 to 2020, respectively. He is currently an Area Editor of the IEEE TRANSACTIONS ON  COMMUNICATIONS.
\end{IEEEbiography}

\begin{IEEEbiography}[{\includegraphics[width=1in,height=1.25in,clip,keepaspectratio]{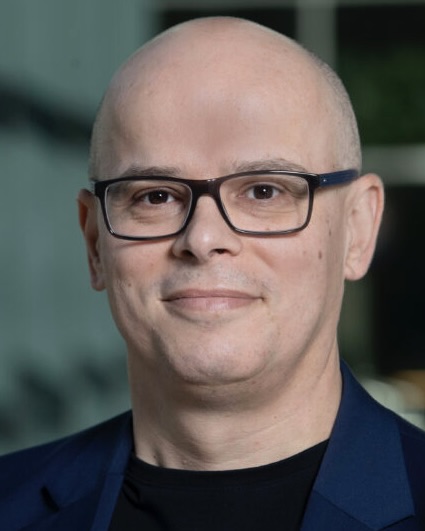}}]{Petar Popovski}
(Fellow, IEEE) is a Professor at Aalborg University, where he heads the section on Connectivity and a Visiting Excellence Chair at the University of Bremen. He received his Dipl.-Ing and M. Sc. degrees in communication engineering from the University of Sts. Cyril and Methodius in Skopje and the Ph.D. degree from Aalborg University in 2005. He received an ERC Consolidator Grant (2015), the Danish Elite Researcher award (2016), IEEE Fred W. Ellersick prize (2016), IEEE Stephen O. Rice prize (2018), Technical Achievement Award from the IEEE Technical Committee on Smart Grid Communications (2019), the Danish Telecommunication Prize (2020) and Villum Investigator Grant (2021). He was a Member at Large at the Board of Governors in IEEE Communication Society 2019-2021. He is currently an Editor-in-Chief of IEEE JOURNAL ON SELECTED AREAS IN COMMUNICATIONS and a Chair of the IEEE Communication Theory Technical Committee. Prof. Popovski was the General Chair for IEEE SmartGridComm 2018 and IEEE Communication Theory Workshop 2019. His research interests are in the area of wireless communication and communication theory. He authored the book ``Wireless Connectivity: An Intuitive and Fundamental Guide’'.
\end{IEEEbiography}

\begin{IEEEbiography}[{\includegraphics[width=1in,height=1.25in,clip,keepaspectratio]{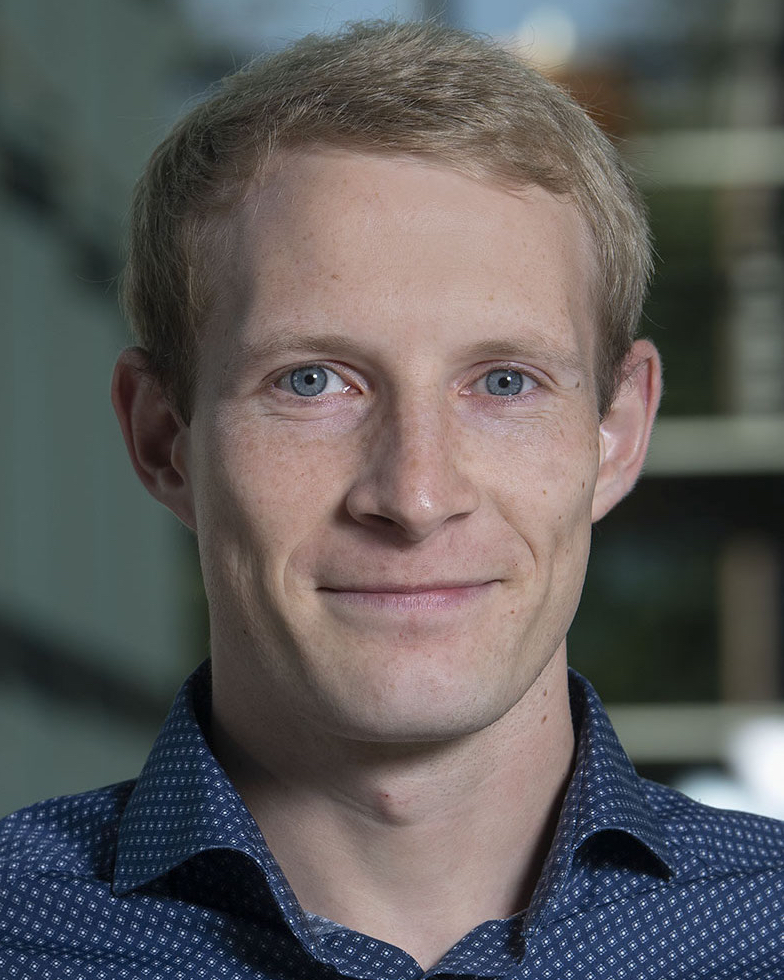}}]{Anders E. Kalør}
(Member, IEEE) received the B.Sc. and M.Sc. degrees in computer engineering and the Ph.D. degree in wireless communications from Aalborg University in 2015, 2017, and 2022, respectively. In 2017, he was a Visiting Researcher with Bosch, Germany, and King’s College London, U.K., in 2020. He is currently a Post-Doctoral
Researcher with The University of Hong Kong, supported by an Individual International Post-Doctoral Grant from the Independent Research Fund Denmark. He is also affiliated with the Connectivity
Section, Aalborg University, Denmark. His current research interests include communication theory and the intersection between wireless communications, machine learning, and data mining for the IoT. He received the Spar Nord Foundation Research Award for his Ph.D. project in 2023.
\end{IEEEbiography}

\begin{IEEEbiography}[{\includegraphics[width=1in,height=1.25in,clip,keepaspectratio]{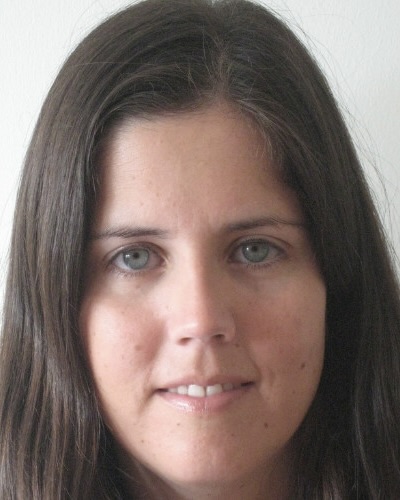}}]{Beatriz Soret}
(Senior Member, IEEE) received the M.Sc. and Ph.D. degrees in telecommunications from the University of Málaga, Spain, in 2002 and 2010, respectively. She has held industrial positions at Nokia Bell Labs and GomSpace. She is currently a Senior Research Fellow with the Telecommunications Research Institute, University of Malaga, and a part-time Associate Professor with Aalborg University. Her current research interests include semantic communications and AoI, LEO satellite communications, and intelligent IoT environments. She received the Best Paper Award from IEEE Globecom in 2013 and the Beatriz Galindo Senior Grant in Spain in 2020.
\end{IEEEbiography}

\vfill
\end{document}